\documentclass[]{aa}
\usepackage{natbib}
\bibpunct{(}{)}{;}{a}{}{,}
\usepackage{graphicx}
\def\vt  {$v_ {t}$}
\begin{document}

\title{First Stars VIII -- Enrichment of the neutron-capture elements
in the early Galaxy. 
\thanks {Based on observations made with the ESO Very Large Telescope 
at Paranal Observatory, Chile (program ID 165.N-0276(A); P.I: R. Cayrel).
}}


\author{
P. Fran\c cois \inst{8,1}, 
E. Depagne \inst{1,10}, 
V. Hill\inst {1}, 
M. Spite\inst {1}, 
F. Spite \inst {1}, 
B. Plez\inst {2}, 
T. C. Beers\inst {3}, 
J. Andersen\inst {5,9}, 
G. James \inst{8,1},
B. Barbuy\inst {4}, 
R. Cayrel\inst {1}, 
P. Bonifacio\inst {1,6},
P. Molaro \inst {6},
B. Nordstr\"om \inst {5} \and 
F. Primas\inst {7}
       } 

\offprints {P. Fran\c cois}

\institute {
      GEPI, Observatoire de Paris-Meudon, CNRS, Univ. de Paris Diderot,
      Place Jules Janssen, F-92190 Meudon, France
         \and
             GRAAL, Universit\'e de Montpellier II, F-34095 Montpellier
             Cedex 05, France
         \and
  Dept. of Physics \& Astronomy, CSCE: Center for the Study of Cosmic Evolution,
  and JINA: Joint Institute for Nuclear Astrophysics, Michigan State University,
            E. Lansing, MI  48824, USA 
         \and
    IAG, Universidade de S\~ao Paulo, Departamento de Astronomia, CP
         3386, 01060-970 S\~ao Paulo, Brazil
         \and
         The Niels Bohr Institute, Astronomy Group, Juliane Maries Vej 30,
         DK-2100 Copenhagen, Denmark 
        \and
    Istituto Nazionale di Astrofisica - Osservatorio Astronomico di Trieste,
    Via G.B. Tiepolo 11, I-34131
             Trieste, Italy
        \and 
         European Southern Observatory (ESO),
         Karl-Schwarschild-Str. 2, D-85749 Garching b. M\"unchen, Germany
	 \and 
	 European Southern Observatory (ESO), 
         Alonso de Cordova 3107, Vitacura,
  	 Casilla 19001,  Santiago 19, Chile
	 \and 
       Nordic Optical Telescope, Apartado 474,
       ES-38700 Santa Cruz de La Palma, Spain
         \and 
       Las Cumbres Observatory, 
       Santa Barbara, California, USA
}

\authorrunning {Fran\c cois et al.}
\mail{Patrick.Francois@obspm.fr}

\titlerunning {First Stars VIII -- Abundances of the neutron-capture elements}

\date{Received 24/04/2007 / accepted 18/09/2007}

\abstract
{Extremely metal-poor (EMP) stars in the halo of the Galaxy are sensitive probes
of the production of the first heavy elements and the efficiency of mixing in
the early interstellar medium. The heaviest measurable elements in such stars are our main guides to understanding the nature and astrophysical site(s) of
early neutron-capture nucleosynthesis.}
{Our aim is to measure accurate, homogeneous neutron-capture element 
abundances for the sample of 32 EMP giant stars studied earlier 
in this series, including 22 stars with [Fe/H] $< -$3.0. } 
{Based on high-resolution, high S/N spectra from the ESO VLT/UVES, 1D, LTE model
atmospheres, and synthetic spectrum fits, we determine abundances or upper limits for the 16 elements Sr, Y, Zr, Ba, La, Ce, Pr, Nd, Sm, Eu, Gd, Dy, Ho, Er, Tm, and Yb in all stars.}
{As found earlier, [Sr/Fe], [Y/Fe], [Zr/Fe] and [Ba/Fe] are below Solar in the EMP stars, with very large scatter. However, we find a tight anti-correlation 
of [Sr/Ba], [Y/Ba], and [Zr/Ba] with [Ba/H] for $-4.5 <$ [Ba/H] $< -2.5$, also 
when subtracting the contribution of the main $r$-process as measured by [Ba/H]. 
Spectra of even higher S/N ratio are needed to confirm and extend these results below [Fe/H] $\simeq -3.5$. The huge, well-characterised scatter of the 
[n-capture/Fe] ratios in our EMP stars is in stark contrast to the negligible dispersion in the [$\alpha$/Fe] and [Fe-peak/Fe] ratios for the same stars found in Paper V. }
{These results demonstrate that a second (``weak'' or LEPP) $r$-process dominates the production of the lighter neutron-capture elements for [Ba/H] $< -2.5$. The  
combination of very consistent [$\alpha$/Fe] and erratic [n-capture/Fe] ratios indicates that inhomogeneous models for the early evolution of the halo are needed. Our accurate data provide strong constraints on future models of the production and mixing of the heavy elements in the early Galaxy.
}

\maketitle
%

\section{Introduction}

In cold dark matter models for hierarchical galaxy formation, the very first generation of metal-free (Population III) stars are thought to be born in 
sub-galactic fragments of mass $M > 5~10^{5} M_{\odot}$ \citep{Ful00,Yos03, 
Mad04}. Recent models of primordial star formation 
\citep{Abe00,Bro05} suggest that these stars were very massive ($M > 100 
M_{\odot}$), although substantial uncertainties remain. 

It is likely that none
of these stars survives in the Galaxy today. However, this first generation 
of stars left imprints of its nucleosynthetic history in the elemental
abundance patterns of the most metal-poor lower-mass stars that we can observe 
at present. Detailed chemical analyses of the most metal-poor stars can 
therefore provide insight into the synthesis of the first heavy elements and how efficiently they were mixed and incorporated in later stellar generations - i.e. how large spiral galaxies such as our own were first assembled.

In Paper V of this series \citep{Cay04}, we confirmed the existence of relatively uniform $\alpha$-element overabundances in 32 very metal-poor 
halo giants down to [Fe/H]$\simeq -4.2$, as expected for material enriched by massive progenitors. The very small dispersion in [$\alpha$/Fe] showed that previous findings of significant scatter in [$\alpha$/Fe] and [Fe-peak/Fe] at low metallicity were due to problems in the data and/or analyses (low S/N, uncertain stellar atmospheric parameters, combinations of data using different line lists, different analysis techniques, etc.). The results of Paper V
thus suggested that mixing of the ISM in the early Galaxy was quite efficient.  

In contrast, the neutron-capture elements have been found to behave very differently \citep{Mol90,Nor93,Pri94}. For example, the [Ba/Fe] and [Sr/Fe] ratios are found to be generally below solar for stars with [Fe/H] 
$< -2.5$ \citep{Mcw95,Rya96,Mcw98}, but the trends with metallicity differ from one element to another. Moreover, several [n-capture/Fe] ratios exhibit a 
large spread at low metallicity \citep{Mcw95,Rya96, Mcw98}, as confirmed recently by \citet{Ce04} and \citet{Bar05} from a large sample of very 
metal-poor stars.

The detailed abundance ratios between the neutron-capture elements are our best diagnostics of the processes that synthesised these elements in the earliest stars. The detailed characteristics of the dispersion of these ratios around the mean relations (amplitude, change with metallicity, etc.) are also our most important diagnostics of the efficiency of mixing in the early ISM. 

As in Paper V, we therefore want to determine, from high-quality spectra analysed in a consistent manner, the precise abundance relations between the main groups of neutron-capture elements seen in the most metal-poor stars and quantify the scatter around these mean relations. For this, we select the same sample of very metal-poor halo giants as discussed earlier in the ``First Stars'' project, using the same spectra, atmospheric parameters, and analysis techniques as before. 

Throughout this paper we will use the designations Very Metal-Poor (VMP), Exrtremely Metal-Poor (EMP), and Ultra Metal-Poor (UMP) for stars with metallicities $-3<$[Fe/H]$<-2$, $-4<$[Fe/H]$<-3$, and [Fe/H]$<-4$, 
respectively \citep{Bee05}.
We will not discuss the Carbon-Enhanced Metal-Poor (CEMP) stars, many of which exhibit peculiar abundances and may be binary systems \citep[and in preparation]{Luc05}.

\begin {table}[ht]
\caption {The observed sample of stars, with adopted model parameters 
(T$_{\rm eff}$, log $g$, \vt, [Fe/H]$_{m}$) and final iron abundances [Fe/H]$_{c}$ (from Paper V).}
\label{tab-parmod}
\begin {center}
\begin {tabular}{lccccc}
Star     & T$_{eff}$ & log g& \vt   &[Fe/H]$_{m}$& [Fe/H]$_{c}$\\
\hline
 HD~2796        & 4950 & 1.5 & 2.1  &$-$2.4& $-$2.47\\
 HD~186478      & 4700 & 1.3 & 2.0  &$-$2.6& $-$2.59\\
 BD~+17:3248    & 5250 & 1.4 & 1.5  &$-$2.0& $-$2.07\\

 BD~--18:5550   & 4750 & 1.4 & 1.8  &$-$3.0& $-$3.06\\
 CD~--38:245    & 4800 & 1.5 & 2.2  &$-$4.0& $-$4.19\\
 BS~16467--062  & 5200 & 2.5 & 1.6  &$-$4.0& $-$3.77\\

 BS~16477--003  & 4900 & 1.7 & 1.8  &$-$3.4& $-$3.36\\
 BS~17569--049  & 4700 & 1.2 & 1.9  &$-$3.0& $-$2.88\\
 CS~22169--035  & 4700 & 1.2 & 2.2  &$-$3.0& $-$3.04\\

 CS~22172--002  & 4800 & 1.3 & 2.2  &$-$4.0& $-$3.86\\
 CS~22186--025  & 4900 & 1.5 & 2.0  &$-$3.0& $-$3.00\\
 CS~22189--009  & 4900 & 1.7 & 1.9  &$-$3.5& $-$3.49\\

 CS~22873--055  & 4550 & 0.7 & 2.2  &$-$3.0& $-$2.99\\
 CS~22873--166  & 4550 & 0.9 & 2.1  &$-$3.0& $-$2.97\\
 CS~22878--101  & 4800 & 1.3 & 2.0  &$-$3.0& $-$3.25\\

 CS~22885--096  & 5050 & 2.6 & 1.8  &$-$4.0& $-$3.78\\
 CS~22891--209  & 4700 & 1.0 & 2.1  &$-$3.0& $-$3.29\\
 CS~22892--052  & 4850 & 1.6 & 1.9  &$-$3.0& $-$3.03\\

 CS~22896--154  & 5250 & 2.7 & 1.2  &$-$2.7& $-$2.69\\
 CS~22897--008  & 4900 & 1.7 & 2.0  &$-$3.5& $-$3.41\\
 CS~22948--066  & 5100 & 1.8 & 2.0  &$-$3.0& $-$3.14\\

 CS~22952--015  & 4800 & 1.3 & 2.1  &$-$3.4& $-$3.43\\
 CS~22953--003  & 5100 & 2.3 & 1.7  &$-$3.0& $-$2.84\\
 CS~22956--050  & 4900 & 1.7 & 1.8  &$-$3.3& $-$3.33\\

 CS~22966--057  & 5300 & 2.2 & 1.4  &$-$2.6& $-$2.62\\
 CS~22968--014  & 4850 & 1.7 & 1.9  &$-$3.5& $-$3.56\\
 CS~29491--053  & 4700 & 1.3 & 2.0  &$-$3.0& $-$3.04\\

 CS~29495--041  & 4800 & 1.5 & 1.8  &$-$2.8& $-$2.82\\
 CS~29502--042  & 5100 & 2.5 & 1.5  &$-$3.0& $-$3.19\\
 CS~29516--024  & 4650 & 1.2 & 1.7  &$-$3.0& $-$3.06\\

 CS~29518--051  & 5200 & 2.6 & 1.4  &$-$2.8& $-$2.69\\
 CS~30325--094  & 4950 & 2.0 & 1.5  &$-$3.4& $-$3.30\\
\hline
\end {tabular}
\end {center}
\end {table}

\section{Observations}

The observations were performed during several observing runs
in 1999 and 2000 at the VLT-Kueyen telescope with the high-resolution
spectrograph UVES \citep{Dek00}. Details of these observations and the
spectrograph settings were given in Paper V, which also provided abundances 
of the lighter elements for the same sample of stars as studied here. 

The spectra were reduced using the UVES package within MIDAS, which performs 
bias and inter-order background subtraction (object and flat-field), optimal extraction of the object (above sky, rejecting cosmic-ray hits), division by 
a flat-field frame extracted with the same weighted profile as
the object, wavelength calibration, rebinning to a constant wavelength step, and
merging of all overlapping orders. The spectra were then added and normalized to
unity in the continuum. 

Because UVES is so efficient in the near UV, we achieve typical S/N ratios per 
resolution element of 50 or more at 350 nm. Thus, the weak lines from the heavy elements become measurable even in the EMP stars of our sample; most previous studies were based on spectra of substantially lower quality. 

\section {Abundance analysis}

As described in Paper V, a classical LTE analysis of our spectra was carried 
out, using OSMARCS model atmospheres \citep{Gus75,Ple92,Edv93,Asp97, Gus03}. 
Abundances were determined with a current version of the {\tt turbospectrum} code \citep{Alv98}, which treats scattering in detail. Solar abundances were adopted from \citet{Gre00}. 

Line detection and equivalent-width measurement was first carried out with the line list of the appendix of Paper I \citep{Hil02} and the automatic code {\tt fitline}, which is based on genetic algorithms. As most of the lines are weak and located in crowded spectral regions, this turned out to be less than optimal, so we decided to determine the abundances by fitting synthetic spectra to all visible lines (and therefore do not list individual measured equivalent widths here).

It soon became clear that establishing upper limits for the abundances of many 
of the heavy elements could also be useful, even when no line from the strongest predicted transition could be detected. These upper limits were computed by comparing the synthetic and observed spectra, and changing the abundance until the computed strength of the line was of the same order as the noise in the observed spectrum. 

All the measured abundances and upper limits are given in Tables \ref{abon1}--\ref{abon3} and shown in detail in Figs.~\ref{SBS1}--\ref{SBS4}.

\subsection{Atmospheric parameters} 

The procedures employed to derive T$_{\rm eff}$, log $g$, and micro-turbulent
velocity estimates \vt for our stars were described in detail in Paper V 
(Sect. 3). In summary, T$_{\rm eff}$ is derived from broadband photometry,
using the \citet{Alo99} calibration. The surface gravity is set by
requiring that the Fe and Ti abundances derived from neutral and singly ionised
transitions be identical. Micro-turbulent velocities are derived by eliminating
the trend in abundance of the Fe~I lines as a function of equivalent width. 
Table~\ref{tab-parmod} lists the atmospheric parameters adopted from Paper V.

\begin {table}[th]
\caption{Estimated errors in the element abundance ratios [X/Fe] and [X/Ba] for BS~17569-049. The other stars yield similar results.}
\label{errors}
\begin {tabular}{lrrr} 
\hline
[X/Fe] & $\Delta$T$_{\rm eff}$ = & $\Delta$log $g$ = & $\Delta v_{t}$ = \\
       &  +100 K &  +0.2 &  +0.2 km s$^{-1}$     \\
\hline 

Sr     &  $-$0.02  &    0.03   &      0.01   \\
Y      &  $-$0.01  &    0.06   &      0.00   \\
Zr     &  $-$0.01  &    0.07   &      0.03   \\
Ba     &     0.02  &    0.03   &   $-$0.02   \\
La     &     0.00  &    0.06   &      0.00   \\
Ce     &     0.00  &    0.07   &      0.05   \\
Pr     &     0.00  &    0.07   &      0.04   \\
Nd     &     0.00  &    0.07   &      0.04   \\
Sm     &     0.00  &    0.07   &      0.05   \\
Eu     &     0.01  &    0.01   &   $-$0.04   \\
Gd     &  $-$0.01  &    0.07   &      0.03   \\
Dy     &     0.00  &    0.07   &      0.04   \\
Ho     &     0.00  &    0.07   &      0.02   \\
Er     &     0.00  &    0.06   &   $-$0.03   \\
Tm     &  $-$0.01  &    0.04   &      0.02   \\
Yb     &     0.01  &    0.06   &   $-$0.01   \\
\hline
[X/Ba] & $\Delta$T$_{\rm eff}$ = & $\Delta$log $g$ = & $\Delta v_{t}$ =\\
       &  +100 K &  +0.2 &  +0.2 km s$^{-1}$     \\
\hline 
Sr     &  $-$0.04  &      0.00   &      0.03   \\
Y      &  $-$0.03  &      0.03   &      0.02   \\
Zr     &  $-$0.03  &      0.04   &      0.05   \\
Ba     &   ...  .  &      ...    &       ...   \\
La     &  $-$0.02  &      0.03   &      0.02   \\
Ce     &  $-$0.02  &      0.04   &      0.07   \\
Pr     &  $-$0.02  &      0.04   &      0.06   \\
Nd     &  $-$0.02  &      0.04   &      0.06   \\
Sm     &  $-$0.02  &      0.04   &      0.07   \\
Eu     &  $-$0.01  &   $-$0.02   &   $-$0.02   \\
Gd     &  $-$0.03  &      0.04   &      0.05   \\
Dy     &  $-$0.02  &      0.04   &      0.06   \\
Ho     &  $-$0.02  &      0.04   &      0.04   \\
Er     &  $-$0.02  &      0.03   &   $-$0.01   \\
Tm     &  $-$0.01  &      0.01   &      0.05   \\
Yb     &  $-$0.01  &   $-$0.03   &      0.01   \\
\hline
\end {tabular}
\end {table}

\subsection{Line list}

For all of the stars in our sample, we adopt the [Fe/H] abundances derived in 
Paper V, which are based on a large number of lines (60--150 Fe~I lines and 
4--18 Fe~II lines). The line list used to determine the heavy-element abundances is taken from Paper I, but updated with recent determinations of oscillator strengths and hyperfine structure corrections \citep{DenH03,Law04}
for several of the elements. 

The solar abundances from \citet{Gre00} have not been corrected for the changes introduced by these corrections, as they are small and only affect some of the transitions in each element.

\subsection {Error budget}

Table~\ref{errors} lists the computed errors in the heavy-element abundance ratios due to typical uncertainties in the stellar parameters. These errors were
estimated by varying T$_{\rm eff}$, log $g$, and $v_{t}$ in the model atmosphere of BS~17569-049 by the amounts indicated; other stars of the sample yield 
similar results. As will be seen, errors in the basic parameters largely cancel out in the abundance ratios between elements in similar stages of ionization and with similar excitation potentials. 

The global error of an element abundance [A/H], including errors in fitting of the synthetic line profile to the observed spectra, is of the order of 
0.20$-$0.25 dex, depending on the species under consideration. The typical 
line-to-line scatter (standard deviation) for a given element is 0.05$-$0.15 
dex.

\begin{figure*}[]
  \centering
  \includegraphics[height=15cm]{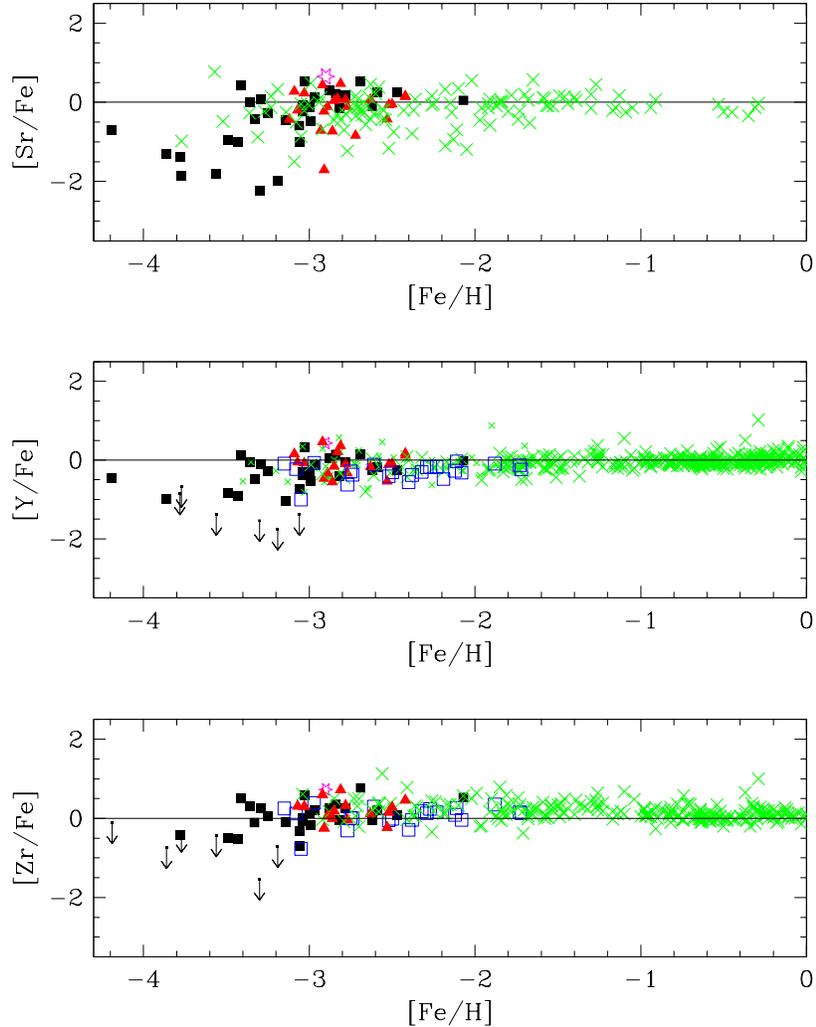}
\caption{The abundance ratios [Sr/Fe], [Y/Fe], and [Zr/Fe] as functions 
of [Fe/H]. {\it Black rectangles:} present study; {\it red triangles}; \citet{Hon04}; {\it blue rectangles}: \citet{Joh02}; {\it green crosses}: selected results from earlier literature (see references listed in text). CS~31082-001 (Paper I) is shown by a {\it magenta star}. }
\label{FIG1}%
\end{figure*}

\begin {table*} 
\caption {Abundance results. Numbers in parenthesis indicate the number of lines measured. } 
\label {abon1}					       
\begin {tabular}{lrrrrrrr} 
\hline
  Object    & [Fe/H]    & [Sr/Fe]       &        [Y/Fe]    &      [Zr/Fe]     &     [Ba/Fe]    &      [La/Fe]     &   [Ce/Fe]        \\
 \hline 
HD~2796      &  $-$2.47 &     +0.25 (3) &     $-$0.25 (10) &      +0.08 (17)  &    $-$0.14 (2) &     $-$0.10 (7)  &       +0.06 (2)  \\
HD~186478    &  $-$2.59 &     +0.24 (3) &     $-$0.19 (9)  &      +0.21 (14)  &    $-$0.04 (4) &       +0.05 (5)  &       +0.01 (15) \\
BD~+17:3248  &  $-$2.07 &     +0.05 (3) &     $-$0.02 (3)  &       +0.53 (5)  &      +0.69 (2) &       +0.66 (5)  &       +0.49 (2)  \\
BD~-18:5550  &  $-$3.06 &   $-$1.01 (2) &  $<$$-$1.38      &    $-$0.71 (5)   &    $-$0.74 (4) & $<$ $-$0.91      & $<$$-$0.42      \\
CD~-38:245   &  $-$4.19 &   $-$0.71 (3) &     $-$0.45 (2)  & $<$$-$0.11       &    $-$0.76 (2) &   $<$ +0.02      & $<$   +0.71      \\
BS~16467-062 &  $-$3.77 &   $-$1.85 (2) &  $<$$-$0.67      & $<$$-$0.33       & $<$$-$1.16     & $<$   +0.30      & $<$   +1.19      \\
BS~16477-003 &  $-$3.36 &     +0.01 (2) &     $-$0.05 (9)  &      +0.31 (13)  &    $-$0.45 (4) & $<$ $-$0.01      & $<$   +0.38      \\
BS~17569-049 &  $-$2.88 &     +0.31 (2) &       +0.04 (10) &      +0.27 (17)  &      +0.20 (4) &       +0.38 (7)  &       +0.23 (9)  \\
CS~22169-035 &  $-$3.04 &   $-$0.08 (2) &     $-$0.38 (10) &    $-$0.08 (15)  &    $-$1.19 (2) & $<$ $-$0.93      & $<$ $-$0.34      \\
CS~22172-002 &  $-$3.86 &   $-$1.31 (2) &     $-$0.98 (1)  & $<$$-$0.74       &    $-$1.17 (2) & $<$ $-$0.01      & $<$   +0.58      \\  
CS~22186-025 &  $-$3.00 &   $-$0.12 (2) &     $-$0.31 (9)  &      +0.14 (15)  &      +0.02 (4) &       +0.19 (6)  &       +0.27 (1)  \\  
CS~22189-009 &  $-$3.49 &   $-$0.95 (2) &     $-$0.83 (1)  &    $-$0.49 (1)   &    $-$1.29 (2) & $<$ $-$0.18      & $<$   +0.41      \\  
CS~22873-055 &  $-$2.99 &   $-$0.48 (2) &     $-$0.53 (9)  &    $-$0.17 (15)  &    $-$0.45 (4) &     $-$0.47 (5)  & $<$ $-$0.09      \\  
CS~22873-166 &  $-$2.97 &     +0.13 (2) &     $-$0.13 (9)  &      +0.20 (15)  &    $-$0.70 (4) &     $-$0.77 (3)  &     $-$0.34 (2)  \\  
CS~22878-101 &  $-$3.25 &   $-$0.27 (2) &     $-$0.28 (9)  &      +0.05 (6)   &    $-$0.58 (2) &     $-$0.42 (1)  & $<$   +0.17      \\  
CS~22885-096 &  $-$3.78 &   $-$1.39 (2) &  $<$$-$0.86      &    $-$0.42 (1)   &    $-$1.10 (1) & $<$ $-$0.09      & $<$   +0.90      \\  
CS~22891-209 &  $-$3.29 &     +0.08 (2) &     $-$0.10 (9)  &      +0.26 (12)  &    $-$0.55 (4) &     $-$0.28 (1)  & $<$ $-$0.19      \\  
CS~22892-052 &  $-$3.03 &     +0.53 (3) &       +0.33 (10) &      +0.58 (17)  &      +1.01 (5) &       +1.11 (6)  &       +1.02 (14) \\  
CS~22896-154 &  $-$2.69 &     +0.54 (2) &       +0.15 (2)  &      +0.77 (4)   &      +0.51 (3) &       +0.42 (1)  &       +0.71 (2)  \\  
CS~22897-008 &  $-$3.41 &     +0.44 (2) &       +0.12 (4)  &      +0.50 (4)   &    $-$1.00 (2) & $<$ $-$0.46      & $<$   +0.33      \\  
CS~22948-066 &  $-$3.14 &   $-$0.46 (2) &     $-$1.05 (2)  &    $-$0.09 (3)   &    $-$0.94 (3) & $<$ $-$0.73      & $<$ $-$0.04      \\  
CS~22952-015 &  $-$3.43 &   $-$0.99 (2) &     $-$0.90 (7)  &    $-$0.52 (3)   &    $-$1.33 (3) & $<$ $-$0.54      & $<$   +0.05      \\  
CS~22953-003 &  $-$2.84 &     +0.22 (3) &       +0.14 (10) &      +0.36 (11)  &      +0.49 (5) &       +0.66 (7)  &       +0.66 (2)  \\  
CS~22956-050 &  $-$3.33 &   $-$0.42 (2) &     $-$0.49 (5)  &    $-$0.11 (2)   &    $-$0.78 (3) & $<$ $-$0.24      & $<$   +0.25      \\  
CS~22966-057 &  $-$2.62 &   $-$0.10 (2) &     $-$0.26 (6)  &    $-$0.04 (3)   &    $-$0.24 (4) &       +0.25 (1)  & $<$   +0.34      \\  
CS~22968-014 &  $-$3.56 &   $-$1.80 (2) &  $<$$-$1.38      & $<$$-$0.44       &    $-$1.77 (1) &     $-$0.11 (1)  & $<$   +0.48      \\  
CS~29491-053 &  $-$3.04 &   $-$0.24 (2) &     $-$0.31 (10) &      +0.02 (9)   &    $-$0.89 (4) & $<$ $-$0.93      & $<$ $-$0.34      \\  
CS~29495-041 &  $-$2.82 &   $-$0.15 (2) &     $-$0.41 (10) &    $-$0.04 (16)  &    $-$0.65 (4) &     $-$0.45 (1)  & $<$ $-$0.16      \\  
CS~29502-042 &  $-$3.19 &   $-$1.98 (2) &  $<$$-$1.75      & $<$$-$0.71       &    $-$1.69 (2) & $<$ $-$0.58      & $<$  +0.51       \\  
CS~29516-024 &  $-$3.06 &   $-$0.59 (2) &     $-$0.74 (7)  &    $-$0.31 (3)   &    $-$0.90 (2) &     $-$0.61 (1)  & $<$ $-$0.32      \\  
CS~29518-051 &  $-$2.78 &     +0.18 (2) &     $-$0.06 (10) &      +0.25 (11)  &    $-$0.45 (2) &     $-$0.49 (1)  & $<$   +0.20      \\  
CS~30325-094 &  $-$3.30 &   $-$2.24 (2) &  $<$$-$1.54      & $<$$-$1.54       &    $-$1.88 (2) & $<$ $-$0.27      &       +0.42 (1)  \\  
\hline 		          			                                                                                
\end {tabular}	          			       
\end {table*}

\begin {table*}
\caption{Abundance results (continued).}
\label {abon2}					       
\begin {tabular}{lrrrrrrr} 
\hline
  Object     &   [Fe/H]&      [Pr/Fe]   &      [Nd/Fe]     &     [Sm/Fe]     &       [Eu/Fe]    &        [Gd/Fe]     &   [Dy/Fe]          \\	   
\hline
HD~2796      & $-$2.47 &      +0.26 (1) &     $-$0.13 (5)  &       +0.06 (1) &        +0.11 (1) &       $-$0.05 (3)  &       +0.01 (5)     \\
HD~186478    & $-$2.59 &      +0.25 (2) &       +0.32 (8)  &       +0.41 (2) &        +0.48 (2) &         +0.41 (2)  &       +0.33 (16)    \\
BD~+17:3248  & $-$2.07 &      +0.66 (1) &       +0.69 (2)  &       +0.74 (2) &        +0.93 (3) &         +0.90 (1)  &       +1.01 (2)     \\
BD~-18:5550  & $-$3.06 &     ...        &     $-$0.46 (2)  &  $<$  +0.05     &      $-$0.20 (1) &       $-$0.26 (1)  &  $<$$-$0.08         \\
CD~-38:245   & $-$4.19 & $<$  +0.83     & $<$   +0.19      & $<$   +1.28     &  $<$   +0.38     &    $<$  +0.77      &  $<$  +0.95         \\
BS~16467-062 & $-$3.77 & $<$  +1.31     & $<$   +0.57      & $<$   +1.46     &  $<$   +0.76     &    $<$  +0.75      &  $<$  +1.43         \\
BS~16477-003 & $-$3.36 & $<$  +0.40     & $<$   +0.36      & $<$   +0.65     &  $<$   +0.25     &    $<$  +0.44      &  $<$  +0.72         \\
BS~17569-049 & $-$2.88 &      +0.75 (3) &       +0.43 (17) &       +0.89 (2) &        +0.72 (1) &         +0.62 (3)  &       +0.59 (9)     \\
CS~22169-035 & $-$3.04 & $<$  +0.18     & $<$ $-$0.76      & $<$ $-$0.07     &  $<$ $-$0.67     &    $<$$-$0.28      &  $<$$-$1.10         \\
CS~22172-002 & $-$3.86 & $<$  +0.50     & $<$ $-$0.04      & $<$   +1.05     &  $<$   +0.05     &         +0.54 (2)  &  $<$$-$0.08         \\
CS~22186-025 & $-$3.00 &      +0.24 (1) &       +0.28 (3)  & $<$   +0.59     &        +0.54 (1) &         +0.68 (1)  &       +0.53 (2)     \\
CS~22189-009 & $-$3.49 & $<$  +0.23     & $<$ $-$0.11      & $<$   +0.88     &  $<$ $-$0.02     &         +1.12 (2)  &  $<$$-$0.35         \\
CS~22873-055 & $-$2.99 &    $-$0.27 (1) &     $-$0.29 (2)  & $<$   +0.18     &      $-$0.17 (1) &       $-$0.03 (2)  &     $-$0.24 (3)     \\
CS~22873-166 & $-$2.97 &    $-$0.04 (1) &     $-$0.25 (3)  & $<$ $-$0.04     &      $-$0.30 (1) &    $<$$-$0.55      &  $<$$-$0.87         \\
CS~22878-101 & $-$3.25 & $<$  +0.09     & $<$ $-$0.45      & $<$   +0.34     &      $-$0.06 (1) &    $<$$-$0.07      &     $-$0.29 (1)     \\
CS~22885-096 & $-$3.78 & $<$  +0.62     & $<$   +0.58      & $<$   +1.27     &  $<$   +0.47     &    $<$  +0.86      &  $<$  +0.24         \\
CS~22891-209 & $-$3.29 &      +0.13 (1) &     $-$0.36 (1)  & $<$   +0.28     &      $-$0.09 (1) &       $-$0.13 (1)  &     $-$0.62 (1)     \\
CS~22892-052 & $-$3.03 &      +0.92 (1) &       +1.15 (22) &       +1.50 (4) &        +1.49 (1) &         +1.45 (8)  &       +1.54 (18)    \\
CS~22896-154 & $-$2.69 &      +0.73 (1) &       +0.67 (2)  &       +0.78 (1) &        +0.86 (1) &         +0.87 (1)  &       +0.97 (2)     \\
CS~22897-008 & $-$3.41 & $<$  +0.50     &       +0.01 (1)  & $<$   +0.50     &  $<$ $-$0.20     &    $<$  +0.29      &  $<$$-$0.63         \\
CS~22948-066 & $-$3.14 & $<$$-$0.02     & $<$ $-$0.56      & $<$ $-$0.26     &  $<$ $-$0.57     &    $<$$-$0.08      &  $<$$-$0.80         \\
CS~22952-015 & $-$3.43 & $<$  +0.07     & $<$ $-$0.37      & $<$ $-$0.37     &  $<$ $-$0.28     &    $<$  +0.01      &  $<$$-$0.71         \\
CS~22953-003 & $-$2.84 &      +0.68 (1) &       +0.72 (2)  &       +0.34 (2) &        +1.05 (1) &         +1.01 (4)  &       +1.04 (12)    \\
CS~22956-050 & $-$3.33 & $<$  +0.37     & $<$ $-$0.17      & $<$   +0.03     &  $<$   +0.02     &    $<$  +0.21      &  $<$$-$0.31         \\
CS~22966-057 & $-$2.62 & $<$  +0.76     &       +0.47 (2)  & $<$   +0.32     &        +0.41 (1) &    $<$  +0.30      &       +0.48 (1)     \\
CS~22968-014 & $-$3.56 & $<$  +0.40     & $<$ $-$0.14      & $<$   +0.16     &  $<$   +0.05     &    $<$  +0.14      &  $<$$-$0.28         \\
CS~29491-053 & $-$3.04 & $<$$-$0.12     &     $-$0.46 (1)  & $<$ $-$0.46     &      $-$0.42 (1) &    $<$$-$0.38      &  $<$$-$0.80         \\
CS~29495-041 & $-$2.82 &    $-$0.14 (1) &       +0.28 (2)  & $<$ $-$0.38     &      $-$0.09 (1) &       $-$0.25 (2)  &  $<$$-$0.62         \\
CS~29502-042 & $-$3.19 & $<$  +0.23     & $<$ $-$0.31      & $<$ $-$0.21     &  $<$ $-$0.22     &    $<$  +0.07      &  $<$  +0.55         \\
CS~29516-024 & $-$3.06 & $<$  +0.10     &     $-$0.44 (1)  & $<$ $-$0.54     &      $-$0.25 (1) &    $<$$-$0.36      &     $-$0.59 (1)     \\
CS~29518-051 & $-$2.78 &      +0.42 (1) &       +0.01 (2)  & $<$ $-$0.12     &  $<$ $-$0.13     &    $<$$-$0.04      &  $<$  +0.26         \\
CS~30325-094 & $-$3.30 & $<$  +0.64 (1) &        0.00 (1)  & $<$    0.49     &  $<$ $-$0.11     &    $<$  +0.18      &  $<$$-$0.34         \\
\hline 
\end {tabular}
\end {table*}

\begin {table*}
\caption{Abundance results (continued).}
\label {abon3}					       
\begin {tabular}{lrrrrr} 
\hline
Object &      [Fe/H]     & [Ho/Fe]         & [Er/Fe]           &      [Tm/Fe]     &       [Yb/Fe]       \\	       
\hline					 
HD~2796      &   $-$2.47 &       +0.01 (1) &        +0.11 (2)  & $<$   +1.26      &        $-$0.13 (1)  \\
HD~186478    &   $-$2.59 &  $<$  +0.43 (1) &        +0.47 (3)  &       +0.75 (1)  &          +0.52 (1)  \\
BD~+17:3248  &   $-$2.07 &       +0.91 (1) &        +1.24 (2)  &            ...   &  ...                \\
BD~-18:5550  &   $-$3.06 &     $-$0.20 (1) &      $-$0.12 (2)  &              ... & ...                 \\
CD~-38:245   &   $-$4.19 &  $<$  +0.53     &  $<$   +0.86      &         ...      & ...                 \\
BS~16467-062 &   $-$3.77 &  $<$  +1.71     &  $<$   +1.14      &         ...      &  ...                \\
BS~16477-003 &   $-$3.36 &  $<$  +0.80     &  $<$   +0.53      &         ...      & ...                 \\
BS~17569-049 &   $-$2.88 &       +0.72 (1) &        +0.55 (4)  &       +0.23 (1)  &        +0.60 (1)    \\
CS~22169-035 &   $-$3.04 &  $<$$-$0.42     &  $<$ $-$0.29      &       +1.20 (1)  &     ...             \\
CS~22172-002 &   $-$3.86 &  $<$  +0.70     &  $<$   +0.53      &         ...      & ...                 \\
CS~22186-025 &   $-$3.00 &       +0.44 (1) &        +0.55 (2)  &       +1.64 (1)  &        +0.12 (1)    \\
CS~22189-009 &   $-$3.49 &  $<$  +0.83     &  $<$   +0.56      &         ...      & ...                 \\
CS~22873-055 &   $-$2.99 &  $<$$-$0.37     &      $-$0.24 (2)  &         ...      & ...                 \\
CS~22873-166 &   $-$2.97 &  $<$$-$0.69     &      $-$0.36 (1)  &              ... & ...                 \\
CS~22878-101 &   $-$3.25 &  $<$  +0.19     &  $<$   +0.02      &         ...      & ...                 \\
CS~22885-096 &   $-$3.78 &  $<$  +0.82     &  $<$   +0.55      &         ...      & ...                 \\
CS~22891-209 &   $-$3.29 &  $<$$-$0.17     &      $-$0.24 (1)  &         ...      & ... $-$0.60 (1)     \\
CS~22892-052 &   $-$3.03 &       +1.59 (1) &        +1.49 (4)  &       +1.59 (5)  &    ...              \\
CS~22896-154 &   $-$2.69 &       +0.88 (1) &        +1.01 (2)  &         ...      & ...                 \\ 
CS~22897-008 &   $-$3.41 &  $<$  +0.45     &  $<$   +0.18      &         ...      & ...                 \\
CS~22948-066 &   $-$3.14 &  $<$$-$0.02     &  $<$ $-$0.09      &         ...      & ...                 \\
CS~22952-015 &   $-$3.43 &  $<$$-$0.13     &  $<$   +0.00      &         ...      & ...                 \\
CS~22953-003 &   $-$2.84 &       +1.18 (1) &        +1.06 (2)  &         ...      & ...   +1.02 (1)     \\
CS~22956-050 &   $-$3.33 &  $<$  +0.87     &  $<$   +0.20      &         ...      & ...                 \\
CS~22966-057 &   $-$2.62 &  $<$  +0.46     &        +0.64 (2)  &         ...      & ...                 \\
CS~22968-014 &   $-$3.56 &  $<$  +0.10     &  $<$   +0.43      &         ...      & ...                 \\
CS~29491-053 &   $-$3.04 &  $<$$-$0.32     &  $<$ $-$0.39      &         ...      & ...                 \\
CS~29495-041 &   $-$2.82 &  $<$$-$0.44     &        +0.04 (1)  &         ...      & ...                 \\
CS~29502-042 &   $-$3.19 &   ...           &  $<$   +0.46      &        +1.83 (1) &    ...              \\
CS~29516-024 &   $-$3.06 &  $<$$-$0.40     &      $-$0.37      &         ...      & ...                 \\
CS~29518-051 &   $-$2.78 &  $<$$-$0.48     &  $<$ $-$0.35      &        +2.00 (1) &      ...            \\
CS~30325-094 &   $-$3.30 &  $<$  +0.34     &  $<$   +0.07      &         ...      & ...                 \\
\hline 						       
\end {tabular}					       
\end {table*}

\section {Abundances of the neutron-capture elements}

\subsection {The light neutron-capture elements Sr, Y, and Zr }

In the Solar System, the abundances of Sr, Y, and Zr are dominated by 
$s$-process production \citep{Arl99}. A small fraction of these elements can 
be produced by the weak $s$-process \citep{Pra90}, but this process is not 
expected to be efficient at the low metallicities observed in our sample. 

Fig. \ref{FIG1} shows the abundance ratios [Sr/Fe], [Y/Fe], and [Zr/Fe] as functions of [Fe/H], as determined here and by \citet{Hon04}, together with data selected from earlier papers \citep{Rya91,Nor01,Gra87,Gil88,Gra88,Gra91,Edv93,Gra94,Mcw95, Car97,Nis97,Mcw98,Ste99,Bur00,Fulb00,Car02,Joh02}. Only results based on 
high-resolution, high-S/N spectroscopy are shown here; thus we do not include the recent lower-S/N data by \citet{Bar05}.

Fig. \ref{FIG1} shows a rather similar behaviour for these three elements, i.e. [X/Fe] $\simeq$0 for stars with [Fe/H] above $\simeq -3.0$. Below this 
metallicity, all the abundance ratios drop below the solar values. In other words, the progressive enrichment in these elements only reaches the solar ratio at about [Fe/H] = $-3.0$.

\begin{figure}
  \centering
  \includegraphics[height=7.5cm,angle=-90]{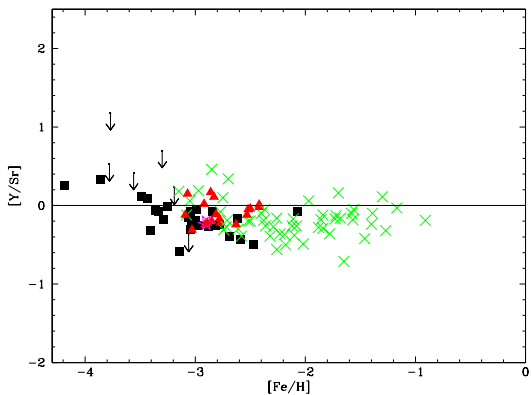}
  \caption{[Y/Sr] as a function of [Fe/H]. Symbols as in Fig.~\ref{FIG1}.  }
  \label{Ysrfe}%
\end{figure}

\subsubsection {Strontium}

Strontium is a key element for probing the early chemical evolution of the
Galaxy, because its resonance lines are strong and can be measured even in 
stars with metallicities as low as [Fe/H] = $-4.0$. For most of our stars, 
only the resonance lines at 4077.719~\AA\ and 4215.519~\AA\ are visible in our spectra. 

We adopt the $~gf$ values from \citet{Sne96} and confirm the large 
underabundance of Sr in EMP stars reported e.g. by \citet{Hon04}. It has long been realized that the [Sr/Fe] ratio exhibits very high dispersion 
for stars with [Fe/H] $ \le -2.8$ \citep{Mcw95,Rya96}, and we confirm this as well. As typical errors in the [Sr/Fe] ratio are no more than a few tenths of a dex at worst, this large spread (over 2 dex) cannot be attributed to observational errors; see, e.g., \citet{Rya96}. 

\begin{figure} 
\centering 
\includegraphics[height=7.5cm,angle=-90]{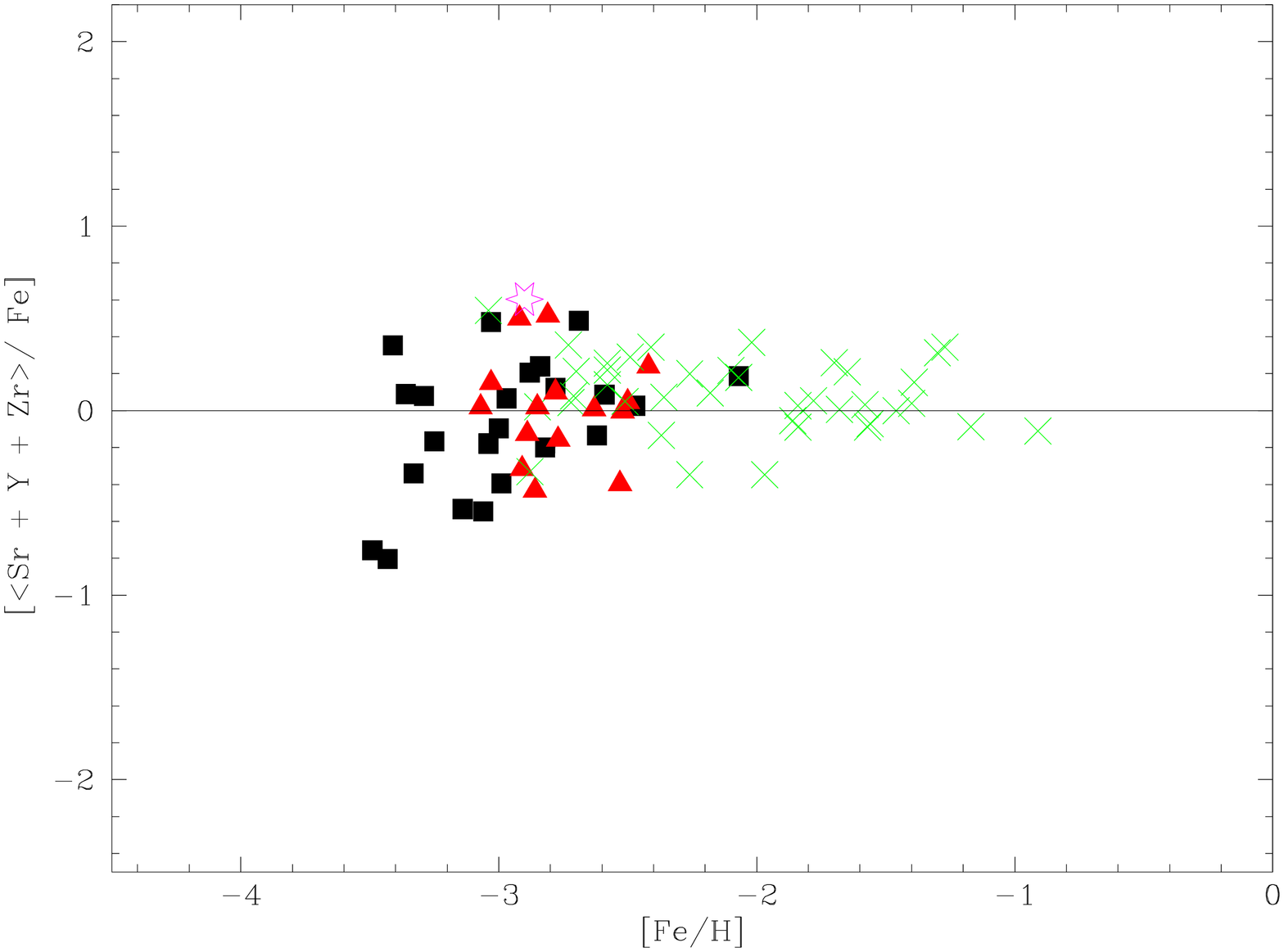}
\caption{[$<$Sr,Y,Zr$>$/Fe] vs. [Fe/H]. Symbols as Fig.~\ref{FIG1}. }
\label{1peakfeb}
\end{figure}

\subsubsection {Yttrium }

The Y lines are somewhat weaker than those of Sr in this temperature range and are not readily detected in our most metal-poor stars. However, nine lines of similar strength (354.9~nm, 360.07~nm, 361.10~nm, 377.43~nm, 378.86~nm, 381.83~nm, 383.29~nm, 395.03~nm, and 439.80~nm) can be measured in stars with [Fe/H]$ > -3.5$ when the temperature is low enough, and then yield rather robust abundance determinations for Y.

The middle panel of Fig.~\ref{FIG1} shows [Y/Fe] as a function of [Fe/H]. The overall trend is similar to that found for Sr, i.e., a solar ratio down to [Fe/H] $\simeq -3.0$, and lower values of increasing dispersion at even lower metallicities.  Unlike Sr, which displays a relatively high dispersion at all metallicities, the weaker and sharper lines of Y yield a very small dispersion in its abundance at intermediate or higher metallicities. The similarity we stress here is that the dispersions in both [Sr/H] and [Y/H] increase by at least a factor of 2 below [Fe/H] $\simeq -3.0$.

\begin{figure}
  \centering
  \includegraphics[height=7.5cm,angle=-90]{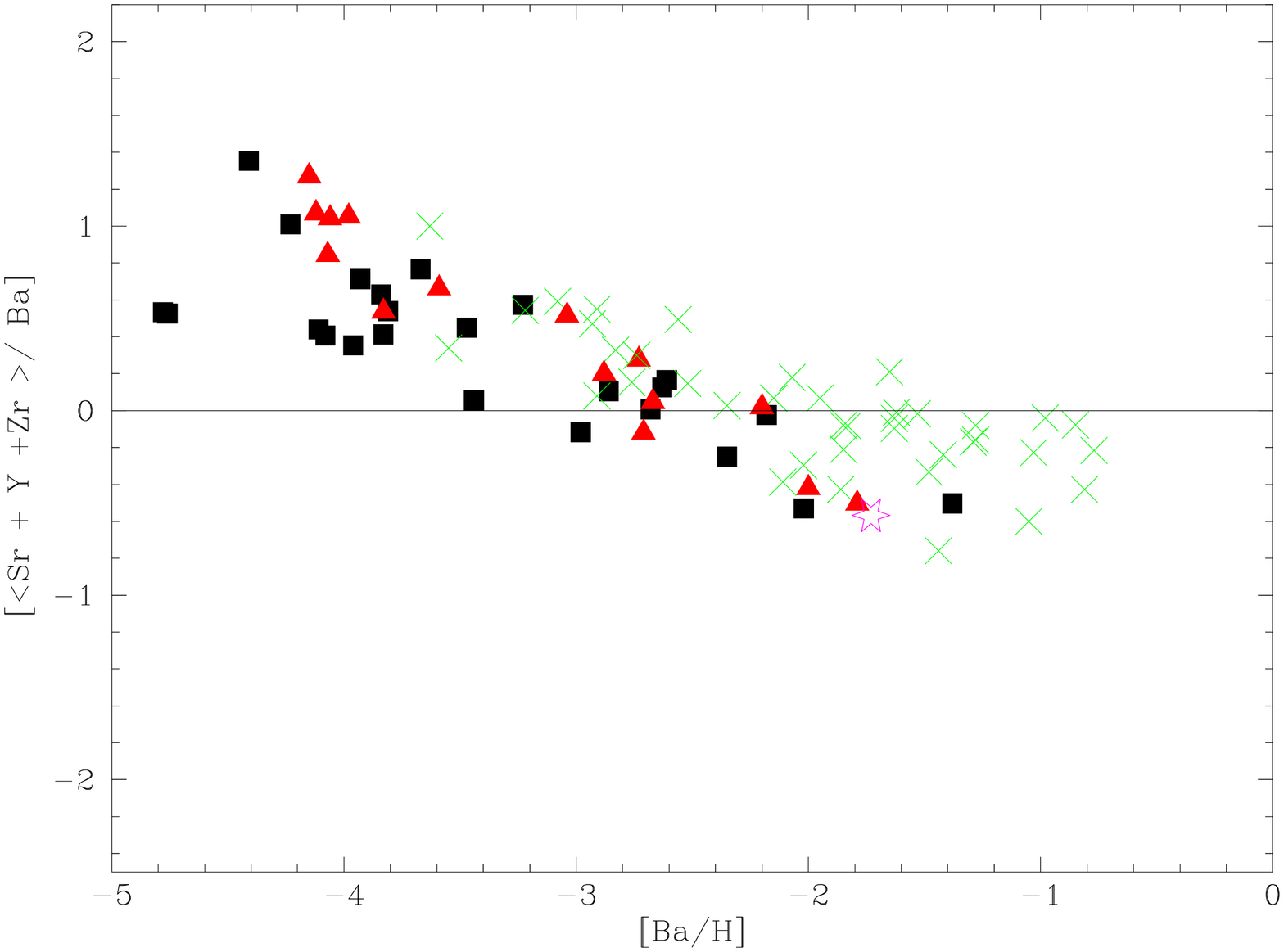}
  \caption{[$<$Sr,Y,Zr$>$/Ba] vs. [Ba/H]. Symbols as Fig.~\ref{FIG1}.  }
  \label{1peakbab}%
\end{figure}

\subsubsection {Zirconium }

Zr is similar to Y in line strength, and we can measure 5-10 lines in stars with [Fe/H] $> -3.50$ (see Fig.~\ref{FIG1}). We find a similar pattern for [Zr/Fe] as for Sr and Y, with a slightly lower average underabundance, large dispersion below [Fe/H] $\simeq -3.0$, and somewhat smaller scatter at intermediate and higher metallicities.

\subsubsection{The [Y/Sr] ratio }

If two elements are formed by the same process, their ratio should not vary 
with metallicity, and the dispersion around the mean value should yield a good estimate of the errors on the abundance determinations. 
Fig.~\ref{Ysrfe} shows the ratio [Y/Sr] as a function of [Fe/H] for our
data, along with those by \citet{Bur00}, \citet{Joh02}, and \citet{Hon04}. 
We confirm that [Y/Sr] is constant with rather low scatter around the mean 
value: [$<$Y/Sr$>$] = $-$0.2$\pm$0.2 (s.d.). This dispersion is fully 
accounted for by the observational errors, indicating that any cosmic scatter 
in this ratio is very small.

\begin{figure}[h]
  \centering
  \includegraphics[height=7.5cm,angle=-90]{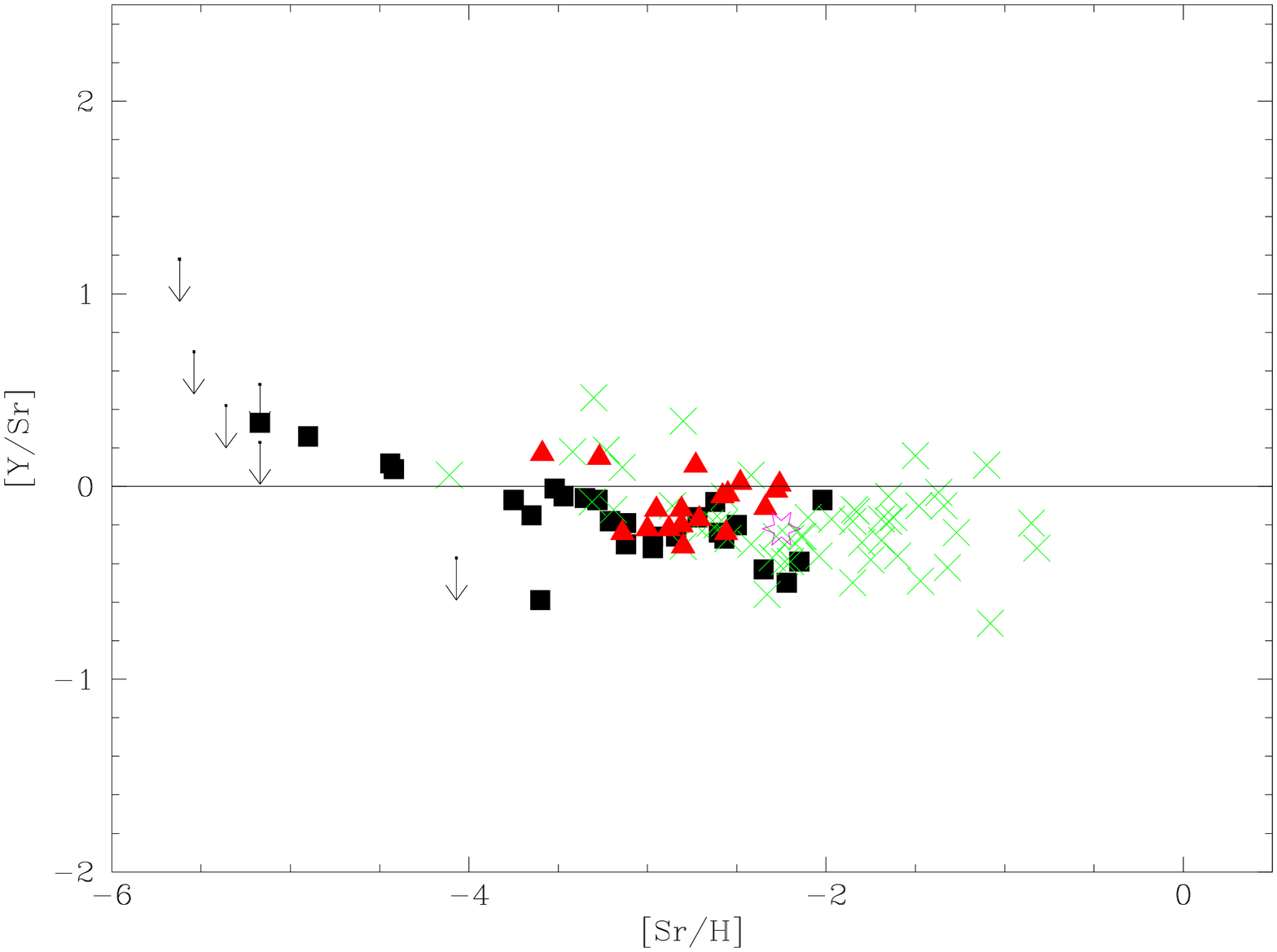}
  \caption{[Y/Sr] vs. [Sr/H]. Symbols as in Fig.~\ref{FIG1}. }
  \label{Ysrsr}%
\end{figure}

However, a plot of [Y/Sr] as a function of [Sr/H] from our data and those by 
\citet{Joh02} (Fig.~\ref{Ysrsr}), appears to show an anticorrelation between [Y/Sr] and [Sr/H] - a result that needs confirmation as some of the data points are upper limits only. This is not seen in the data set of \citet{Hon04}, as 
the range of Sr abundances in their sample is fairly small.

  \begin{figure*}
    \centering
    \includegraphics[height=20cm]{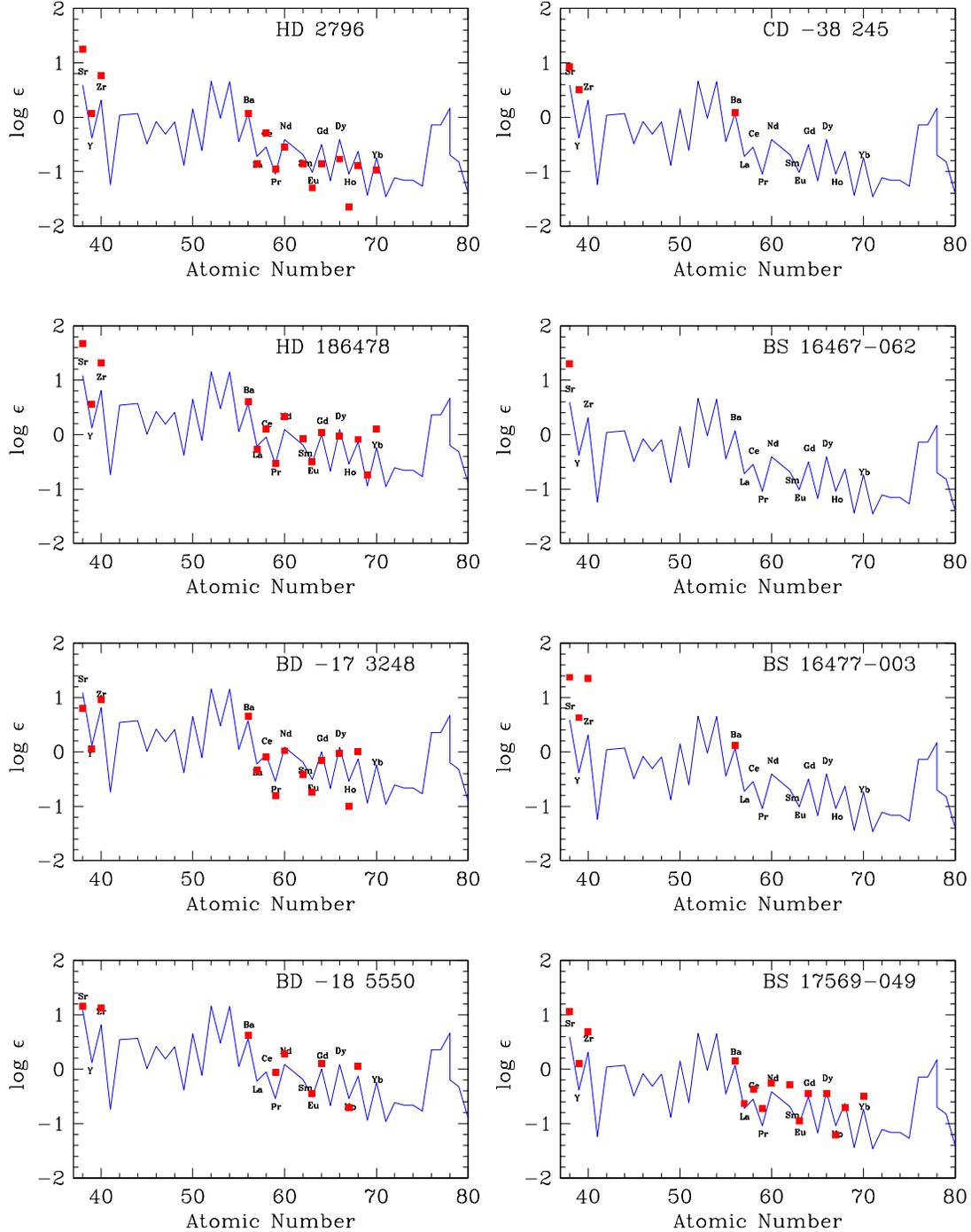}
    \caption{Abundance patterns for the neutron-capture elements in 
     our sample. The full line shows the Solar-system $r$-process
     abundance pattern from \citet{Arl99}, scaled to match the observed
     abundance of Ba in each star.}
    \label{SBS1}%
   \end{figure*}

  \begin{figure*}
    \centering
    \includegraphics[height=20cm]{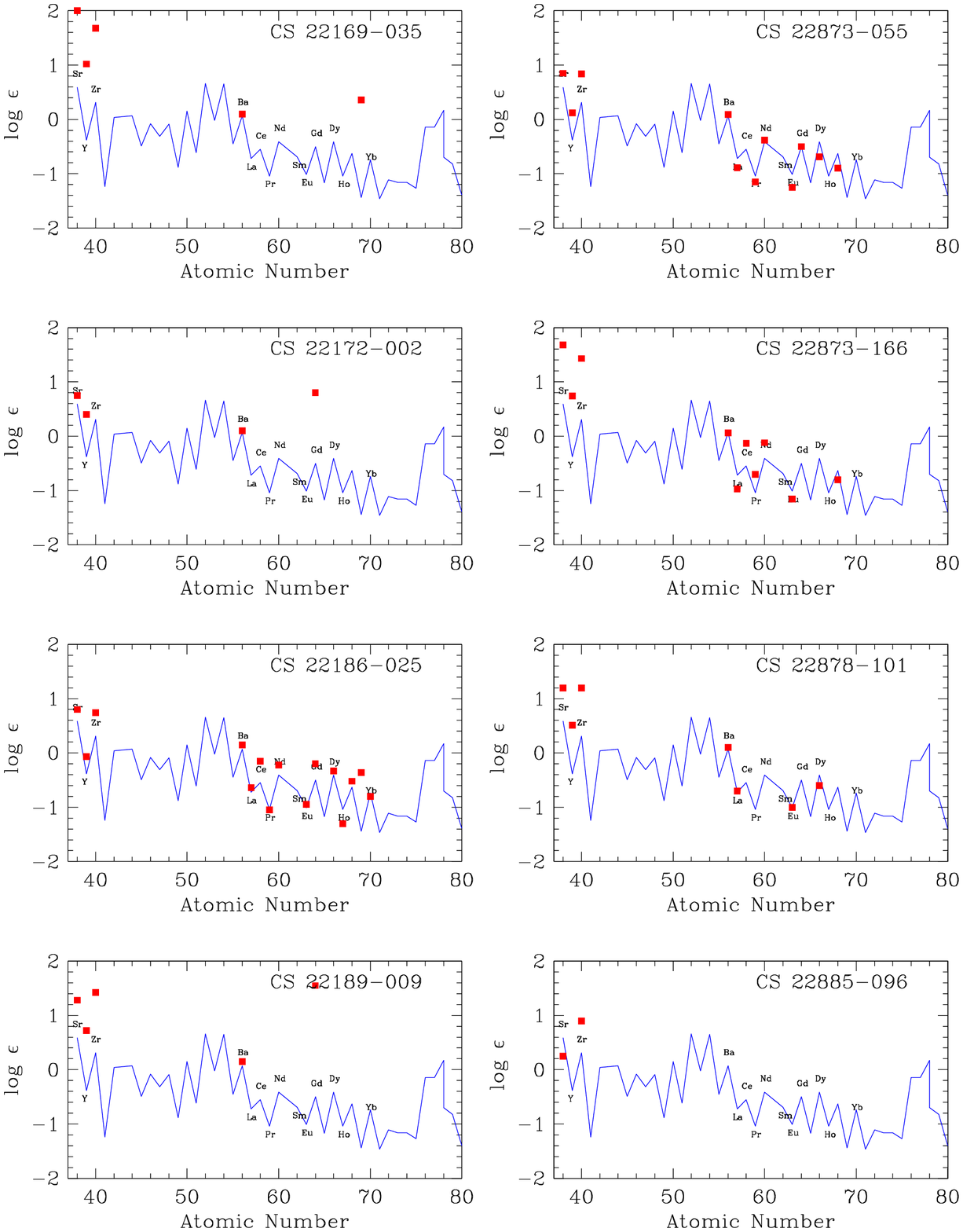}
    \caption{As Fig. \ref{SBS1}, for the next 8 stars. }
    \label{SBS2}%
   \end{figure*}

  \begin{figure*}
    \centering
    \includegraphics[height=20cm]{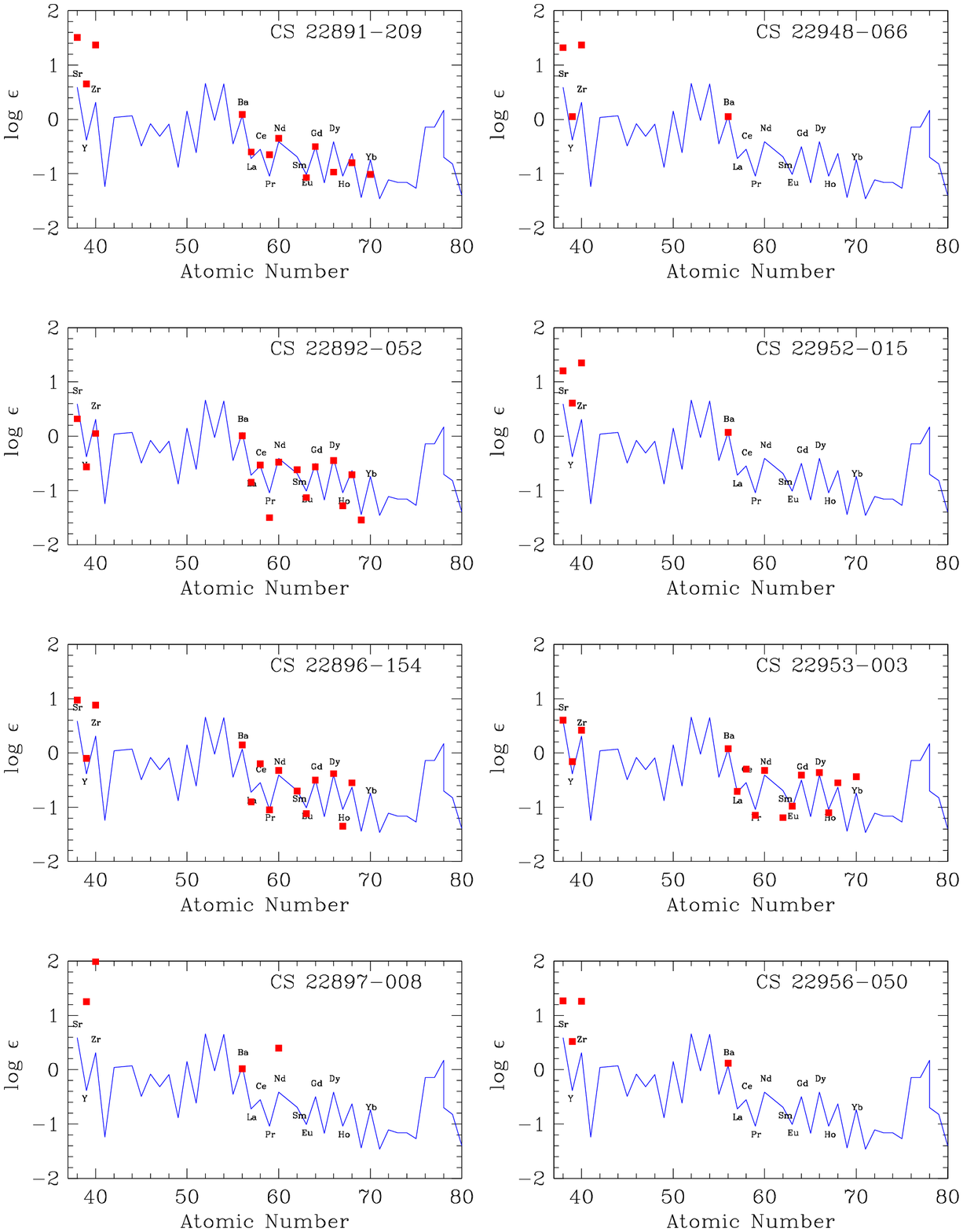}
    \caption{As Fig. \ref{SBS1}, for the next 8 stars.}
    \label{SBS3}%
   \end{figure*}

  \begin{figure*}
    \centering
    \includegraphics[height=20cm]{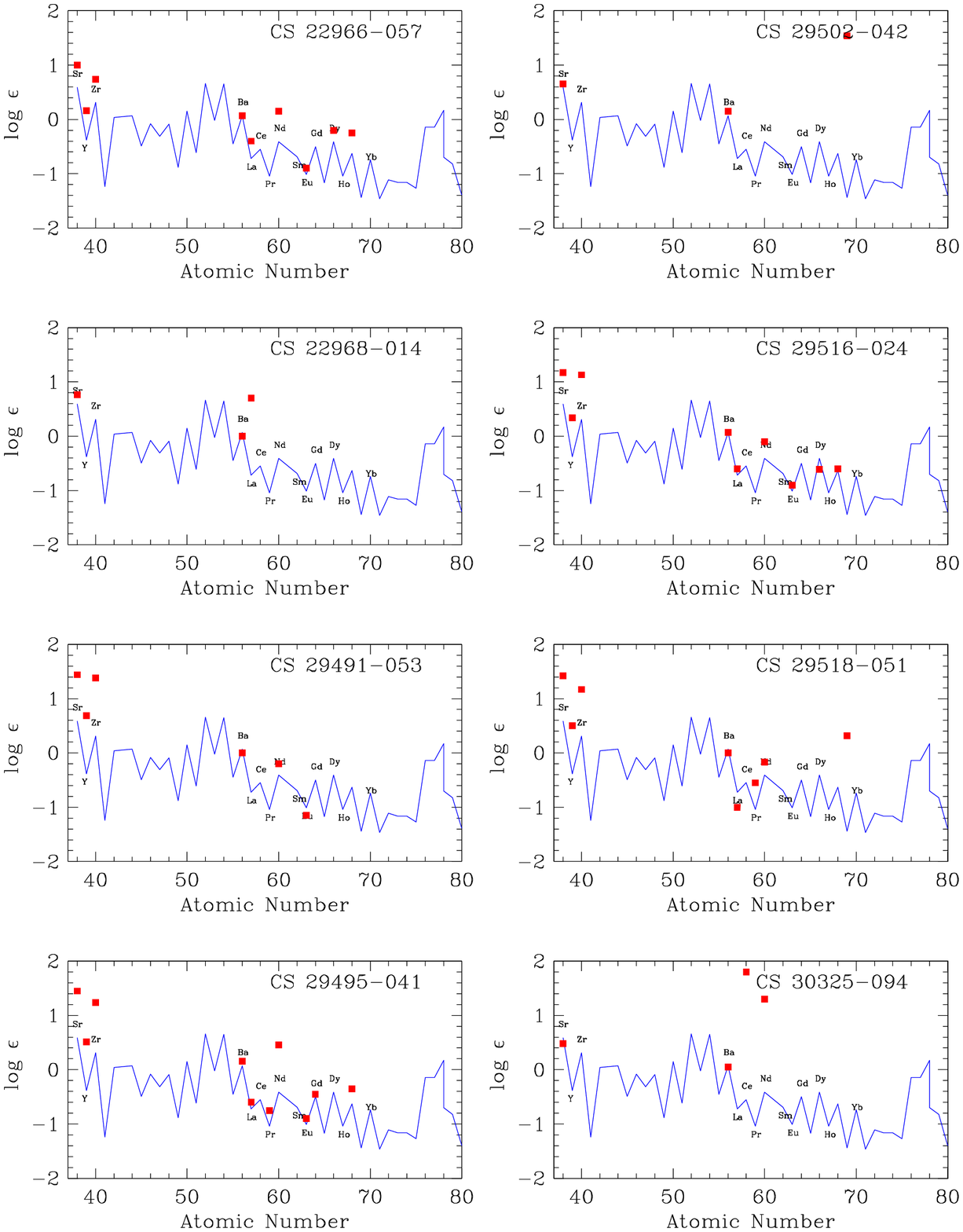}
    \caption{As Fig. \ref{SBS1}, for the last 8 stars.}
    \label{SBS4}%
   \end{figure*}

\subsubsection {General features of the first neutron-capture peak elements}

The first-peak elements are known to have a more complex origin than the heavier neutron-capture elements like Ba or Eu, which are only produced by the ``main'' components of the $r$- and $s$-processes. In solar-type material, Sr, Y and Zr are formed in the ``main'' $s$-process, but at lower metallicity the ``weak'' $s$-process \citep{Bus99} also contributes. In our EMP stars, we expect a pure $r$-process origin for the neutron-capture elements, and we wish to explore the nature of those processes in more detail.

Fig.~\ref{1peakfeb} shows the average [$<$Sr+Y+Zr$>$/Fe] ratio for our stars and from recent literature as a function of [Fe/H]. Only stars with data for all three elements are included, which limits the sample towards the lowest metallicities). We find a clear increase in the dispersion of this ratio with decreasing metallicity. Note also that the two most metal-poor stars ($\rm [Fe/H] \simeq -3.5$) in Fig.~\ref{1peakfeb} are nearly one dex below the solar value, reflecting the strong deficiency of all three elements in the most metal-poor stars.

Because Fe and the neutron-capture elements form under quite different conditions, it may be more informative to study their abundances as functions of another heavy element. The strong resonance lines of Ba can be measured in stars down to almost [Fe/H] $= -4.0$, so we select Ba as our alternative reference element. Fig.~\ref{1peakbab} shows the mean [$<$Sr+Y+Zr$>$/Ba] ratio as a function of [Ba/H]. We find a striking, tight anti-correlation, especially for stars below [Ba/H] $\simeq -2.5$), which may indicate that another nucleosynthesis process produces the light neutron-capture elements preferentially at low metallicity. We discuss this point more fully in Sect. \ref{LEPP}. 

\begin{figure}[ht]
  \centering
  \includegraphics[height=12cm]{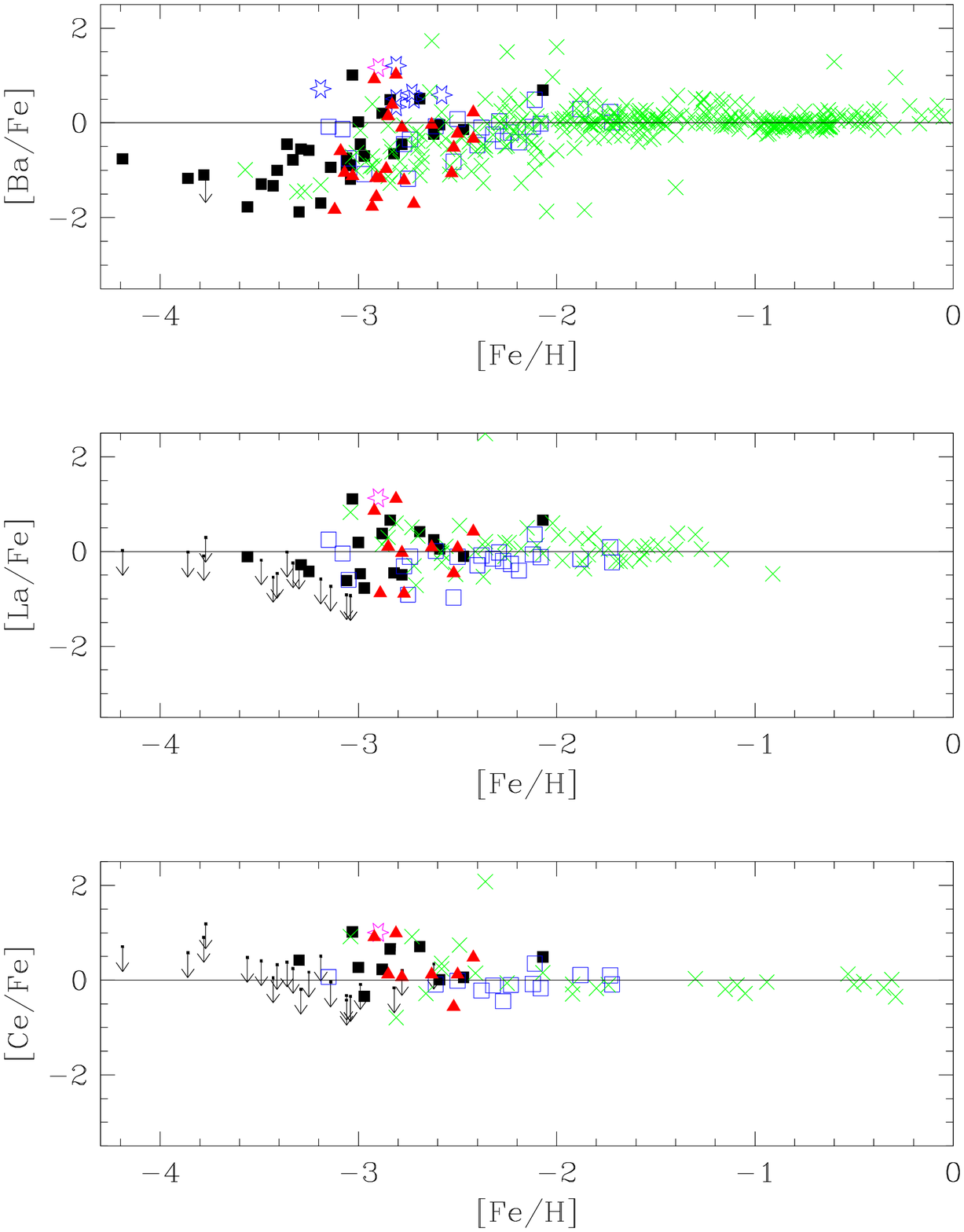}
\caption{[Ba/Fe], [La/Fe], and [Ce/Fe] as functions of [Fe/H]. Symbols as in Fig.~\ref{FIG1}. Blue stars in the upper panel: Data from \citet{Bar05}. }
  \label{FIG2}%
\end{figure}

\subsection {The second neutron-capture peak elements ($56\leq Z\leq 72$)}

This range in atomic mass includes the well-studied elements Ba, Eu, and La. 
Ba and Eu played a key role in understanding early nucleosynthesis, when \citet{Tru81} first suggested that the [Ba/Eu] vs. [Fe/H] observations of \citet{Spi78} could be naturally understood if both of these neutron-capture 
elements were synthesised by the $r$-process in massive stars during early Galactic evolution (85\% of the Ba in the Solar System is due to the 
$s$-process). 

\begin{figure}[ht]
  \centering
  \includegraphics[height=12cm]{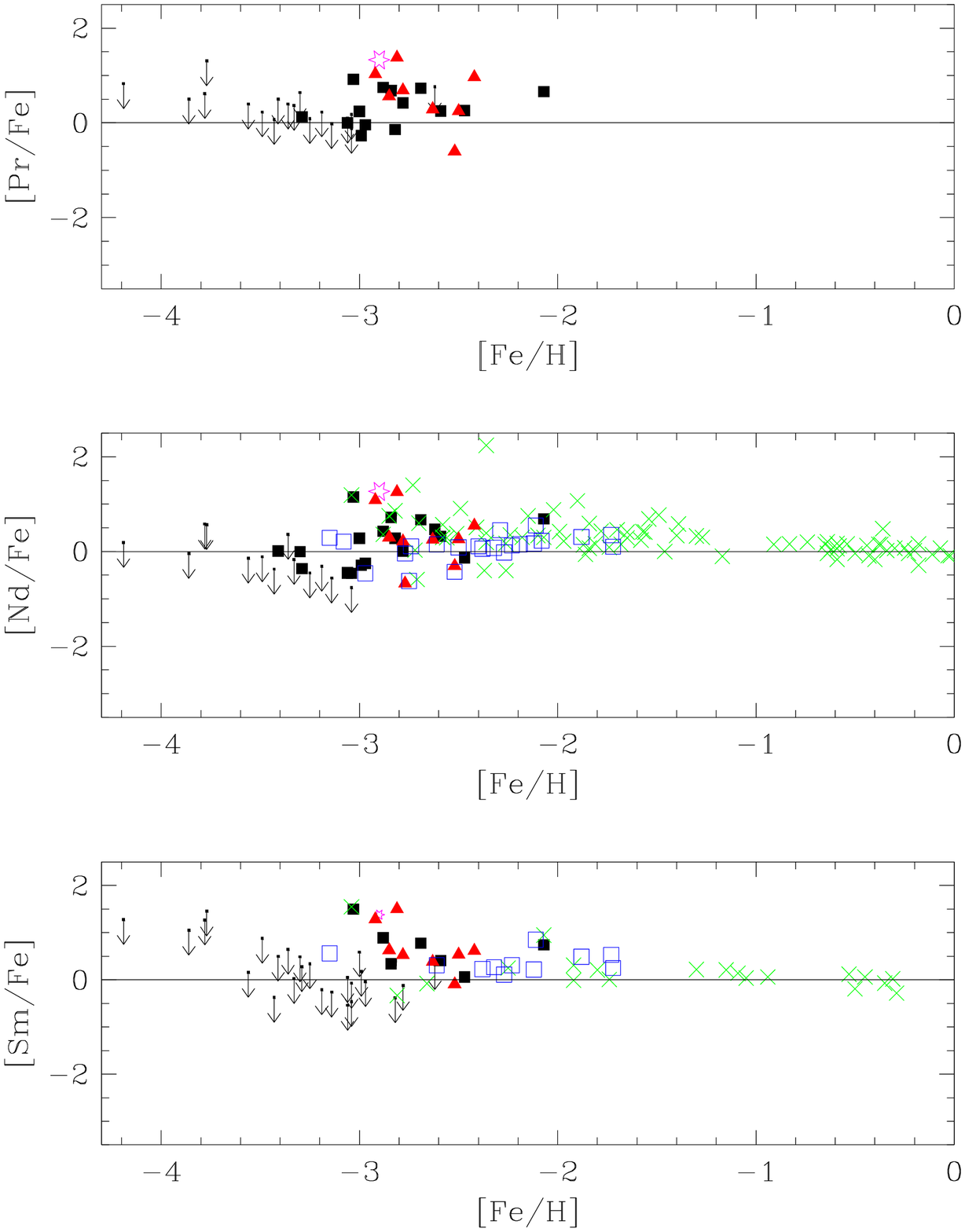}
  \caption{[Pr/Fe], [Nd/Fe], and [Sm/Fe] as functions of [Fe/H]. 
   Symbols as in Fig.~\ref{FIG1}.  }
  \label{FIG3}%
\end{figure}

\begin{figure}[ht]
  \centering
  \includegraphics[height=12cm]{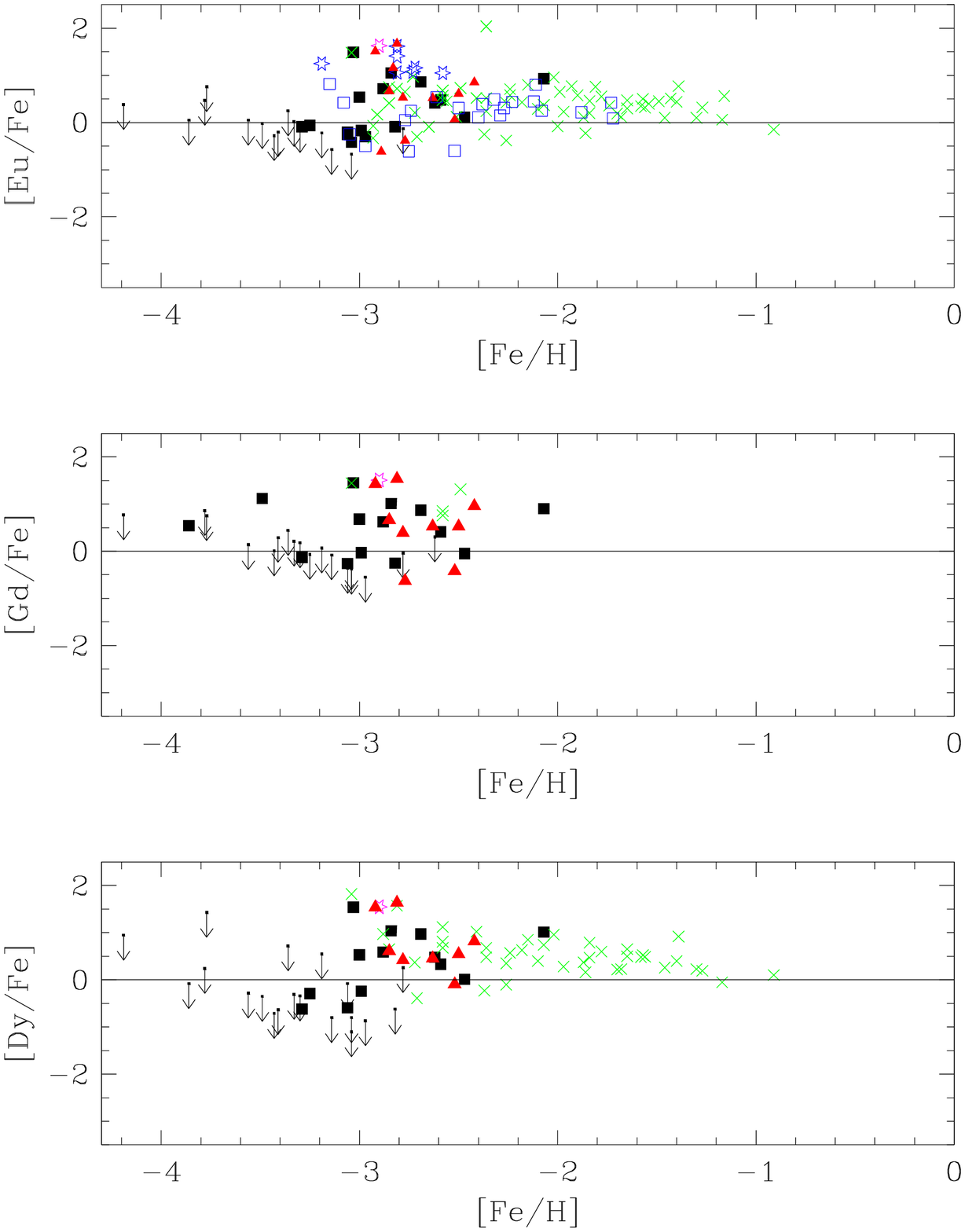}
  \caption{[Eu/Fe], [Gd/Fe], and [Dy/Fe] as functions of [Fe/H]. 
Symbols as in Fig.~\ref{FIG1}. Blue stars in the upper panel: Data from \citet{Bar05}.}
  \label{FIG4}%
\end{figure}

\begin{figure}[ht]
  \centering
  \includegraphics[height=12cm]{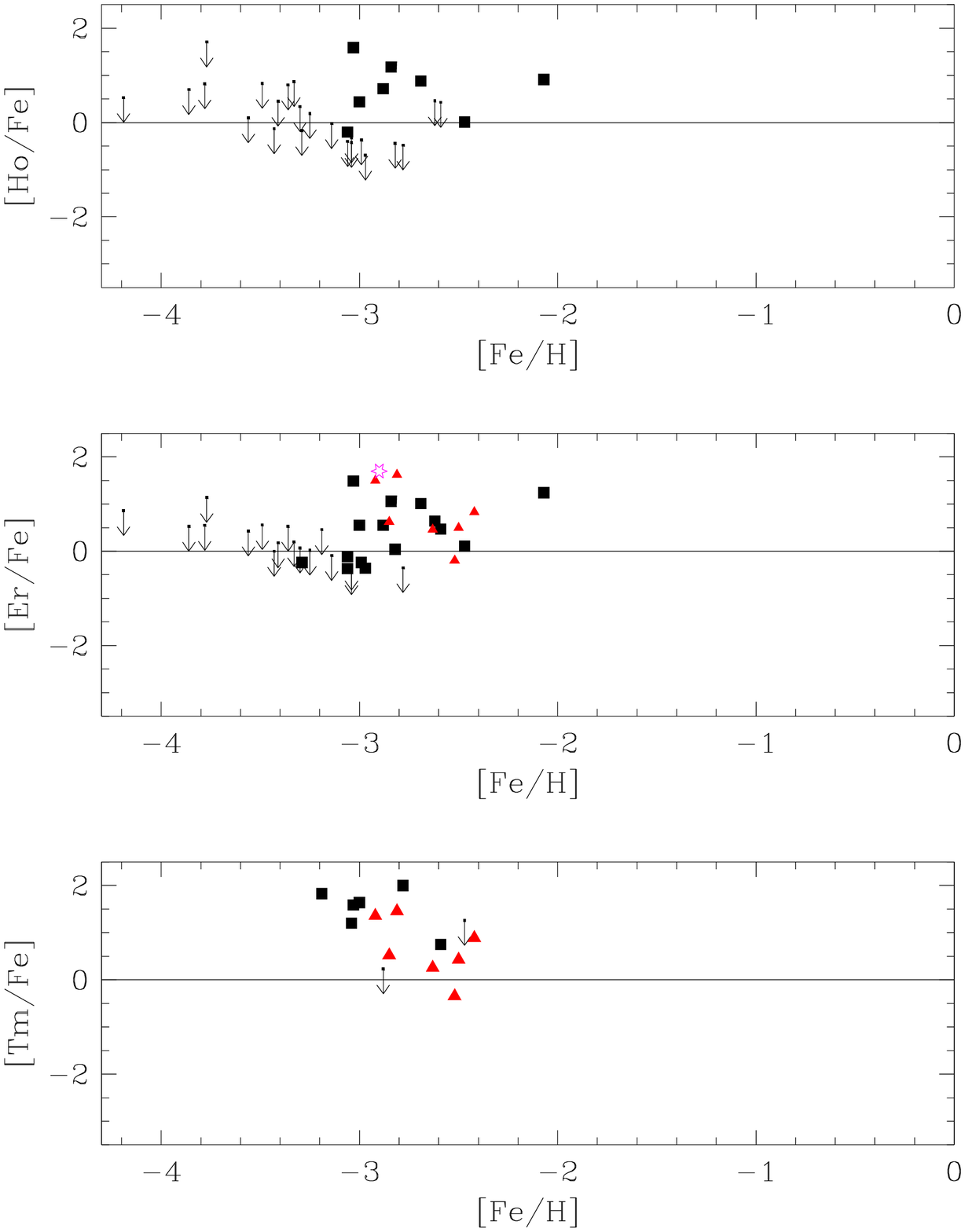}
  \caption{[Ho/Fe], [Er/Fe], and [Tm/Fe] as functions of [Fe/H]. 
 Symbols as in Fig.~\ref{FIG1}. }
  \label{FIG5}%
\end{figure}

Due to the high UV efficiency of UVES, we have been able to determine abundances or upper limits in many of our stars for several other heavy neutron-capture elements (Ce, Pr, Nd, Sm, Gd, Dy, Ho, Er and Tm). The results are shown in Figs. \ref{FIG2} - \ref{FIG5} as functions of [Fe/H], together with those by \citet{Joh02}, \citet{Hon04}, and data selected from earlier literature. These data enable us to discuss the nature of the early $r$-process nucleosynthesis in considerable detail.

As noted above, Ba is a particularly interesting element, in part because
the resonance lines are strong enough to be measured in all but two of our
stars and permit us to explore mean trends and scatter amongst the 
neutron-capture elements down to [Fe/H] = $-4.2$; see Fig.~\ref{FIG2}. All
our abundance results for Ba have been derived assuming the isotopic composition corresponding to the $r$-process \citep{Mcw98}.

In the metallicity range $-$2.5 to $-$3.0, we confirm the very large dispersion in [Ba/Fe] at given [Fe/H] noted by several previous authors, increasing towards the lowest metallicities. Our study adds a significant number of stars below [Fe/H] = $-3.0$. Although the number of stars in this range remains small, Fig.~\ref{FIG2} suggests that [Ba/Fe] continues to decline to a mean value of [Ba/Fe] = $-2.0 - -1.0$, with a declining scatter as well. This might indicate that the nucleosynthesis processes involved undergo significant changes below [Fe/H] = $-3.2$.

The greatest scatter in [Ba/Fe] (a factor of 1000) occurs in the metallicity range $-3.2\le{\rm [Fe/H]}\le-2.8$. Thus, if the Ba and Fe in these stars was created by the same class of progenitor objects, their yields would have to
vary by a similarly large factor, whatever model of chemical evolution for the
early Galaxy one adopts. The yield of Ba could be extremely 
metallicity-dependent or, perhaps more likely, the early production of Ba and 
Fe was not correlated with each other, and Ba and Fe were produced in different
astrophysical sites, as suggested by \citet[ and references therein]{Wan01}.

As Fig.~\ref{FIG2} shows, we do not observe a single star with a [Ba/Fe] ratio above solar below [Fe/H] $\simeq -3.2$; however, we do note that \citet{Bar05} do detect at least a few stars with [Ba/Fe] above solar at metallicities down to [Fe/H] $\simeq -3.4$. 

The metallicity interval showing the largest scatter in [Ba/Fe] ($-3.2\le{\rm [Fe/H]}\le -2.8$) is also where the extremely $r$-process-enhanced metal-poor stars are found; i.e. those with [r-element/Fe] $>$ +1.0, referred to as r-II stars by \citet{Bee05}. CS~22892-052, CS~31082-001, the eight new r-II stars found by \citet{Bar05}, and the most recent discovery HE~1523-0909 \citep{Fre07}, all fall in this range. It is interesting that both CS~22892-052 and CS~31082-001 fit into the same region of Fig.~\ref{FIG2} as the ``normal'' 
(non-r-II) stars (albeit at the very upper limit), so these extreme r-II stars are not exceptional as far as the [Ba/Fe] ratio is concerned. 

Like Ba, both La and Ce are primarily due to the $s$-process at solar metallicity.
For La and Ce, we can determine abundances for stars with [Fe/H] $> -3.0$, but only upper limits for the more metal-poor stars. It is interesting, however, that we find the same increase of the scatter with declining metallicity in the range $-3.2<$ [Fe/H] $<-2.0$ for La and Ce as for Ba. As this is also seen in the data from \citet{Hon04} and earlier literature (see Fig.~\ref{FIG2}), there is little doubt as to its reality.     
 
Figure ~\ref{FIG3} shows our results for [Pr/Fe], [Nd/Fe], and [Sm/Fe], which in Solar-system material are formed by the $s$- and $r$-process in roughly equal proportions. For Pr, the only earlier data are from \citet{Hon04}. We confirm the high [Pr/Fe] ratios found by these authors down to [Fe/H] $\simeq -3.0$. Our upper limits show that a rather large scatter in [Pr/Fe] exists down to [Fe/H] $\simeq -3.0$; for Nd and Sm, the scatter clearly increases with declining metallicity until its maximum at [Fe/H] $\simeq -3.0$.  Note that we have Nd measurements for three stars with [Fe/H] $< -3.2$.

Eu, Gd, and Dy are elements that are produced primarily by the $r$-process, also in Solar-system material (93\%, 84\%, and 87\%, respectively, according to \citet{Arl99}). Figure~\ref{FIG4} shows that they behave similarly to the other elements of the second neutron-capture peak and display increasing overabundances with declining metallicity, accompanied by increasing scatter. Once again, it appears that the scatter is at maximum at [Fe/H] $\simeq -3.0$, as found by \citet{Hon04}. 

Note that for [Eu/Fe], low values $(\leq$0.0) are found only below [Fe/H] $\simeq -3.0$. \citet{Bar05} did find stars with high [Eu/Fe] ($>$ 0.5)
at metallicities lower than [Fe/H] = $-3.0$, but from a much larger sample of stars than ours. This indicates that stars with high [Eu/Fe] ratios are quite rare at very low metallicity, so that dedicated surveys are needed to uncover additional examples. For Gd, we do measure high [Gd/Fe] values in two stars 
(CS~22172-002 and CS~22189-009) below [Fe/H] = $-$3.4.

Finally, Figure~\ref{FIG5} shows our results for Ho, Er, and Tm, also produced almost exclusively in the $r$-process. Very few previous results exist for these three elements, which we find to be generally overabundant, as also reported by \citet{Hon04} for Er and Tm. Once more, the large scatter in the element ratios appears maximal at [Fe/H] = $-3.0$. Our few results for Yb (not plotted) follow the same general trend.

\section {Discussion}

Our accurate, detailed, and homogeneous abundance data for the neutron-capture elements in a large sample of VMP and EMP stars enables us to address two important questions regarding the first stages of heavy-element enrichment in the Galaxy: {\it (i):} The nucleosynthesis process(es) that formed the first heavy elements, and {\it (ii):} the efficiency with which the newly synthesised elements were incorporated in the next generation(s) of stars, including those that have survived until today. We discuss each of these in turn in the following.

\begin{figure}[th]
  \centering
  \includegraphics[height=12cm]{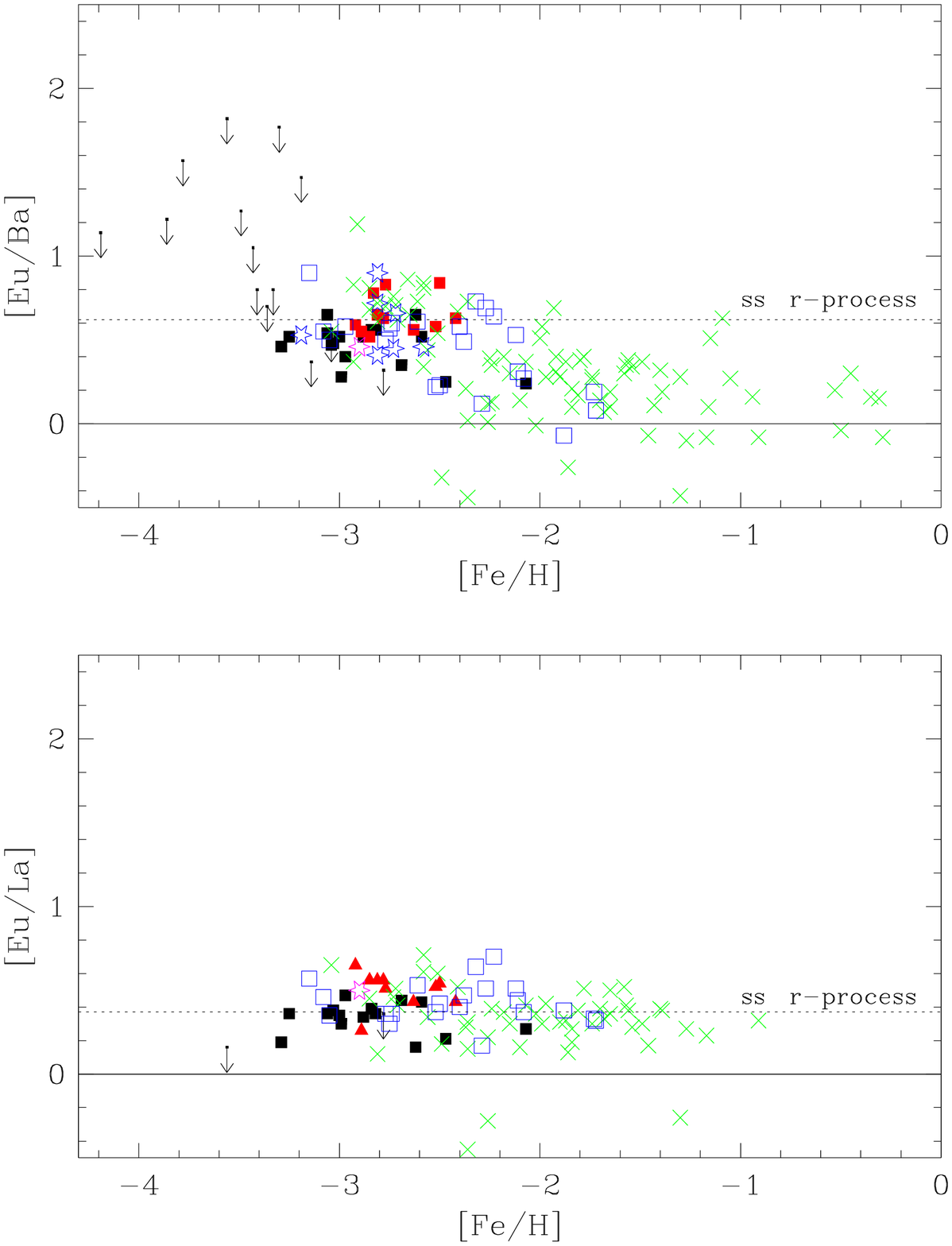}
  \caption{[Eu/Ba] and [Eu/La] as function of [Fe/H]; symbols as in previous
   figures. The dashed lines indicate the Solar-system $r$-process abundance
   ratios \citep{Arl99}}
  \label{FIG6}%
\end{figure}

\subsection {Diagnostics of the $r$-process(es) in EMP stars}

We begin by repeating the classical \citet{Tru81} test of the relative weight of the $r$- and $s$-process as a function of metallicity. Ba and La are produced mostly by the ``main'' $s$-process in Solar-metallicity stars (92\% and 83\%, respectively, according to \cite{Arl99}, but in EMP stars they should be due to the $r$-process. Fig.~\ref{FIG6} shows the [Eu/Ba] and [Eu/La] ratios as a function of [Fe/H] for our stars, along with earlier data. The dashed lines in both panels indicate the Solar-system $r$-process abundance ratios \citep{Arl99}. 

Our [Eu/Ba] ratios do cluster around the Solar-system $r$-process value at low metallicity, but a substantial scatter remains. Some of this may be due to the Ba data because of the broad hyperfine structure of the Ba lines: If the mix of Ba isotopes in the star is different from that assumed in the synthetic spectrum, the fit to the observed spectrum may be less stable than for single-component lines. Indeed, the [Eu/La] ratios exhibit substantially smaller dispersion at all metallicities, demonstrating that the scatter in [Eu/Ba] is essentially due to the Ba, not the Eu abundances. Together, the two panels of Fig.~\ref{FIG6} confirm that the neutron-capture elements in EMP stars were produced predominantly or exclusively by the 
$r$-process.

\begin{figure}[th]
  \centering
  \includegraphics[height=12cm]{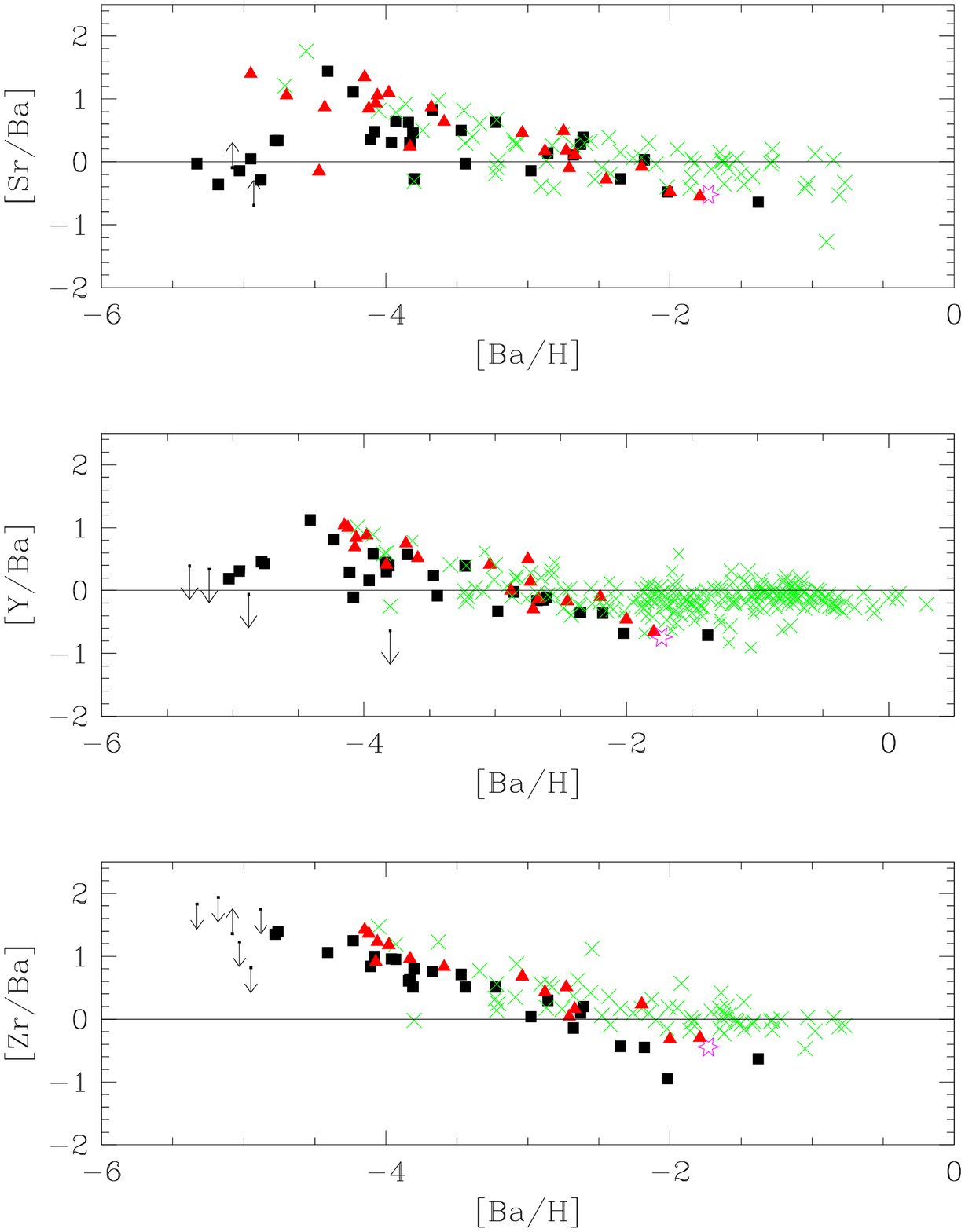}
 \caption{[Sr/Ba], [Y/Ba], and [Zr/Ba] vs. [Ba/H]. 
  Symbols as in Fig.~\ref{FIG1}.}
  \label{SrBa}%
\end{figure}

Given the large scatter of the [n-capture element/Fe] ratios as functions of [Fe/H] (Figs.~\ref{FIG1} and \ref{FIG2} -- \ref{FIG5}), we proceed to compare elemental abundances within the neutron-capture group itself in the following. As noted earlier, we choose Ba as the reference element because data are available for nearly all our stars. 

Fig.~\ref{SrBa} shows the [Sr/Ba], [Y/Ba], and [Zr/Ba] ratios vs. [Ba/H] as determined by us and previous authors. We find a tight anti-correlation of
[X/Ba] with [Ba/H] for all three elements, at least down to [Ba/H] = -4.5. 
We emphasize that most stars in our sample are {\it not} enriched in $r$-process elements, but note that the two extreme r-II stars CS~22892-052 and CS~31082-001 do in fact follow the same relation as the ``normal'' stars. In particular, we find no stars that are both Sr-poor and Ba-rich, as suspected already by \citet{Hon04}; however, such cases {\it are} found among the C-enhanced metal-poor (CEMP) stars \citep{Siv06}. 

Our most Ba-poor stars, below [Ba/H] $\simeq -4.5$, seem to depart from the correlation and show roughly Solar values for [Sr/Ba] and [Y/Ba], although we note that \citet{Hon04} do find a couple of high [Sr/Ba] ratios in this region. This might indicate that the additional production channel for Sr, Y, and Ba discussed below may not operate in the very first stellar generations. However, the sample is very small (these are among our most metal-poor stars, with [Fe/H] $< -3.2$), and more reliable measurements of Sr, Y, and Zr in stars with low [Ba/H] will be needed for a definitive conclusion.

\subsection{Synthesis of the first-peak elements}\label{LEPP}

The diagrams discussed above amply demonstrate that not all the neutron-capture elements in metal-poor stars were produced by a single $r$-process, as discussed by \citet[and references therein]{Tra04}; an additional process must contribute preferentially to the production of the first-peak elements in VMP/EMP stars, previously called the ``weak'' $r$-process; we will discuss below the aptness of this term.

\citet{Tra04} explored the issue by following the Galactic enrichment of Sr, Y, and Zr using homogeneous chemical evolution models. They confirmed that a process of primary nature ($r$-process) is required to explain the observed abundance trends, argued that massive stars were the likely sites as these elements occur at very low metallicity, and coined the term ``Lighter-Element Primary Process'' (LEPP) for it. However, regardless of nomenclature, the actual process, site, or progenitor stars have not been identified.

\citet{Ces05} came to similar conclusions, based on the behavior of Ba and Eu. They confirmed the need for a primary source to explain the behaviour of [Ba/Fe] vs. [Fe/H] and suggested that the primary production of Eu and Ba is associated with stars in the mass range 10-30 M$_{\odot}$. \citet{Ish04} computed the evolution of [Eu/Fe], using inhomogeneous chemical evolution models with induced star formation, and concluded that the observations implied that the 
low-mass range of supernovae were the dominant source of Eu. 

The observations shown in Fig. \ref{SrBa} clearly cannot be explained by single 
$r$-process. The trends suggest the existence of three different regimes: {\it (i):} [Ba/H] $\geq -2.5$, where all ratios are close to Solar; {\it (ii):} $-4.5 \leq$ [Ba/H] $\leq -2.5$, where Sr, Y, and Zr become increasingly overabundant relative to Ba at lower metallicities, and {\it (iii):} [Ba/H] $\leq -4.5$, where the abundance ratios seem to drop to Solar again. The latter transition corresponds to [Fe/H]$\simeq -3$, i.e. the metallicity range in which {\it all} the highly $r$-process enriched metal-poor stars are have been found so far -- the r-II stars as defined by \citet{Bee05}. 

It appears from these plots, and from the great uniformity of the $r$-process element patterns in the r-II stars observed so far, that the main $r$-process dominates the total abundance pattern of the heavy elements once they have been enriched beyond the level of [Ba/H] $\geq -2.5$. At levels up to 2 dex below this threshold, another process contributes increasingly to the production of the first-peak elements Sr, Y, and Zr. We want to clarify the properties of this process as independently of the main $r$-process as possible.

To do so, we have computed the mean residuals of Sr, Y, and Zr in each of our stars from the Solar-system $r$-process abundance pattern of \citet{Arl99} as shown in Figs.~\ref{SBS1}--\ref{SBS4}. Thus, these abundance residuals should represent the pure production of the unknown process, free of interference from the main $r$-process. 
\begin{figure}
  \centering
  \includegraphics[height=7.5cm,angle=-90]{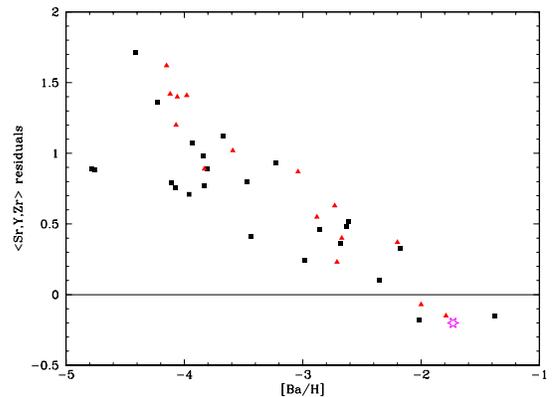}
  \caption{Average abundance residuals of Sr, Y, and Zr from the Solar-system
   curves in Figs.~\ref{SBS1}--\ref{SBS4} vs. overall heavy-element 
   content as measured by [Ba/H]. Symbols as Fig.~\ref{FIG1}.  }
  \label{SSres}%
\end{figure}

The result is shown in Fig. \ref{SSres} and shows that, far from being ``weak'', the LEPP is responsible for 90-95\% of the total abundance of these elements at [Ba/H] $\simeq -4.3$, where the [$<$Sr,Y,Zr$>$/Ba] ratio may split into two branches, as suggested on theoretical grounds by \citet{Ish99} and \citet{Ish00}. 

One would surmise that qualitative differences in neutron exposure or the nature of the available seed nuclei in the most extreme metal-poor stars could cause such differences. E.g., \citet{Qia07} propose that the first-peak elements (Sr, Y, Zr) are formed by charged-particle reactions in the so-called $\alpha$-process \citep{Woos92} in all supernovae, while heavy $r$-process elements would form only in low-mass SNe with O-Ne-Mg cores and iron only in high-mass SNe. The correlation shown in Fig. \ref{SSres} would appear difficult to reconcile with this otherwise interesting scenario.

As an alternative, a new nucleosynthesis process (the $\nu p$-process) has been proposed very recently by \citet{Fro06}. This process should occur in 
core-collapse supernovae and would allow for the nucleosynthesis of nuclei with mass number A $>$ 64.

\subsection{Heavy-element enrichment in the early Galaxy}

\begin {table*}[ht]
\caption {Robust estimates of scatter for the observed abundance ratios (see text). ``[l/Fe]'' represents $[\rm <Sr,Y,Zr>/Fe]$. }
\label {tab-scatter}
\begin {center}
\begin{tabular}{c@{ }c@{ }c@{ }c@{ } c@{~~~ }c@{ }c@{ }c@{ }c@{ }c@{ }c@{ }}
Ord.          &Abs.       & N      &       Range    &  S$_{BI}$ (CL,~CU)            &   Ord.         &Abs.       & N      &        Range   &  S$_{BI}$ (CL,~CU)           \\                                            \\
\hline
$\rm [Sr/Fe]$ &$\rm[Fe/H]$&  31    &     $\leq -$2.0   &   0.433 (0.324,0.542)    &  $\rm [Sm/Fe]$ &$\rm[Fe/H]$&   8    &     $\leq -$2.0   &   0.464 (0.293,0.634)   \\
              &           &  11    &   $-$2.0 to  $-$3.0  &   0.244 (0.166,0.323)    &                &           &        &                &                         \\
              &           &  20    &     $\leq -$3.0   &   0.639 (0.439,0.838)    &  $\rm [Eu/Fe]$ &$\rm[Fe/H]$&  18    &     $\leq -$2.0   &   0.306 (0.177,0.435)   \\
              &           &        &                &                          &                &           &  10    &    $-$2.0 to  $-$3.0 &   0.338 (0.250,0.426)   \\
$\rm [Y/Fe]$  &$\rm[Fe/H]$&  26    &     $\leq -$2.0   &   0.307 (0.259,0.354)    &                &           &   8    &     $\leq -$3.0   &   0.274 (0.000,0.551)   \\
              &           &  11    &   $-$2.0 to  $-$3.0  &   0.199 (0.155,0.243)    &                &           &        &                &                         \\
              &           &  15    &     $\leq -$3.0   &   0.388 (0.309,0.468)    &  $\rm [Gd/Fe]$ &$\rm[Fe/H]$&  14    &     $\leq -$2.0   &   0.526 (0.407,0.646)   \\
              &           &        &                &                          &                &           &   8    &    $-$2.0 to  $-$3.0 &   0.449 (0.275,0.623)   \\
$\rm [Zr/Fe]$ &$\rm[Fe/H]$&  26    &     $\leq -$2.0   &   0.284 (0.212,0.356)    &                &           &   6    &     $\leq -$3.0   &   0.664 (0.431,0.896)   \\
              &           &  11    &   $-$2.0 to  $-$3.0  &   0.244 (0.166,0.323)    &                &           &        &                &                         \\
              &           &  15    &     $\leq -$3.0   &   0.374 (0.254,0.494)    &  $\rm [Dy/Fe]$ &$\rm[Fe/H]$&  13    &     $\leq -$2.0   &   0.202 (0.052,0.353)   \\
              &           &        &                &                          &                &           &   8    &    $-$2.0 to  $-$3.0 &   0.145 (0.095,0.196)   \\
$\rm [Y/Sr]$  &$\rm[Fe/H]$&  25    &     $\leq -$2.0   &   0.104 (0.074,0.133)    &                &           &   5    &     $\leq -$3.0   &   0.743 (0.357,1.130)   \\
              &           &  11    &   $-$2.0 to  $-$3.0  &   0.122 (0.090,0.154)    &                &           &        &                &                         \\
              &           &  14    &     $\leq -$3.0   &   0.083 (0.035,0.131)    &  $\rm [Ho/Fe]$ &$\rm[Fe/H]$&   9    &     $\leq -$2.0   &   0.574 (0.375,0.774)   \\
              &           &        &                &                          &                &           &        &                &                         \\
$\rm [Y/Sr]$  &$\rm[Sr/H]$&  25    &     $\leq -$2.0   &   0.065 (0.051,0.079)    &  $\rm [Er/Fe]$ &$\rm[Fe/H]$&  16    &     $\leq -$2.0   &   0.299 (0.154,0.444)   \\
              &           &  12    &   $-$2.0 to  $-$3.0  &   0.095 (0.049,0.141)    &                &           &  10    &    $-$2.0 to  $-$3.0 &   0.239 (0.130,0.348)   \\
              &           &  13    &     $\leq -$3.0   &   0.053 (0.037,0.070)    &                &           &   6    &     $\leq -$3.0   &   0.584 (0.241,0.926)   \\
              &           &        &                &                          &                &           &        &                &                         \\
$\rm [l/Fe]$  &$\rm[Fe/H]$&  24    &     $\leq -$2.0   &   0.270 (0.201,0.338)    &  $\rm [Tm/Fe]$ &$\rm[Fe/H]$&   6    &     $\leq -$2.0   &   0.398 (0.197,0.598)   \\
              &           &  11    &   $-$2.0 to  $-$3.0  &   0.222 (0.156,0.287)    &                &           &        &                &                         \\
              &           &  13    &     $\leq -$3.0   &   0.363 (0.234,0.492)    &  $\rm [Yb/Fe]$ &$\rm[Fe/H]$&   6    &     $\leq -$2.0   &   0.556 (0.370,0.743)   \\
              &           &        &                &                          &                &           &        &                &                         \\
$\rm [l/Ba]$  &$\rm[Ba/H]$&  24    &     $\leq -$1.0   &   0.232 (0.191,0.272)    &  $\rm [Eu/Ba]$ &$\rm[Fe/H]$&  18    &     $\leq -$2.0   &   0.103 (0.081,0.124)   \\
              &           &   9    &   $-$2.0 to -4.0  &   0.198 (0.131,0.265)    &                &           &  10    &   $-$2.0 to  $-$3.0  &   0.121 (0.084,0.158)   \\
              &           &  15    &     $\leq -$4.0   &   0.264 (0.201,0.328)    &                &           &   8    &     $\leq -$3.0   &   0.078 (0.048,0.109)   \\
              &           &        &                &                          &                &           &        &                &                         \\
$\rm [Ba/Fe]$ &$\rm[Fe/H]$&  30    &     $\leq -$2.0   &   0.412 (0.324,0.501)    &  $\rm [Eu/La]$ &$\rm[Fe/H]$&  15    &     $\leq -$2.0   &   0.037 (0.011,0.064)   \\
              &           &  11    &    $-$2.0 to  $-$3.0 &   0.413 (0.296,0.529)    &                &           &  10    &   $-$2.0 to  $-$3.0  &   0.045 (0.009,0.081)   \\
              &           &  19    &     $\leq -$3.0   &   0.441 (0.286,0.597)    &                &           &  13    &     $\leq -$3.0   &   0.026 (0.000,0.057)   \\
              &           &        &                &                          &                &           &        &                &                         \\
$\rm [La/Fe]$ &$\rm[Fe/H]$&  18    &     $\leq -$2.0   &   0.407 (0.255,0.559)    &  $\rm [Sr/Ba]$ &$\rm[Ba/H]$&  29    &     $\leq -$1.0   &   0.312 (0.264,0.361)   \\
              &           &  11    &    $-$2.0 to  $-$3.0 &   0.409 (0.301,0.517)    &                &           &  19    &   $-$2.0 to -4.0  &   0.290 (0.240,0.340)   \\
              &           &   7    &     $\leq -$3.0   &   0.175 (0.000,0.507)    &                &           &  10    &     $\leq -$4.0   &   0.379 (0.274,0.484)   \\
              &           &        &                &                          &                &           &        &                &                         \\
$\rm [Ce/Fe]$ &$\rm[Fe/H]$&  13    &     $\leq -$2.0   &   0.038 (0.000,0.217)    &  $\rm [Y/Ba]$  &$\rm[Ba/H]$&  26    &     $\leq -$1.0   &   0.200 (0.143,0.256)   \\
              &           &   9    &    $-$2.0 to  $-$3.0 &   0.014 (0.000,0.257)    &                &           &  18    &   $-$2.0 to -4.0  &   0.179 (0.129,0.220)   \\
              &           &   4    &     $\leq -$3.0   &   0.068 (0.000,0.187)    &                &           &   8    &     $\leq -$4.0   &   0.358 (0.211,0.506)   \\
              &           &        &                &                          &                &           &        &                &                         \\
$\rm [Pr/Fe]$ &$\rm[Fe/H]$&  14    &     $\leq -$2.0   &   0.285 (0.162,0.409)    &  $\rm [Zr/Ba]$ &$\rm[Ba/H]$&  25    &     $\leq -$1.0   &   0.221 (0.170,0.272)   \\
              &           &  10    &    $-$2.0 to  $-$3.0 &   0.255 (0.169,0.340)    &                &           &  19    &   $-$2.0 to -4.0  &   0.226 (0.177,0.276)   \\
              &           &   4    &     $\leq -$3.0   &   0.168 (0.000,0.362)    &                &           &   6    &     $\leq -$4.0   &   0.124 (0.000,0.266)   \\
              &           &        &                &                          &  \\
$\rm [Nd/Fe]$ &$\rm[Fe/H]$&  18    &    $\leq -$2.0    &   0.225 (0.131,0.319)    &  \\
              &           &  11    &    $-$2.0 to  $-$3.0 &   0.229 (0.159,0.299)    &  \\
              &           &   7    &     $\leq -$3.0   &   0.135 (0.000,0.376)    &  \\
              &           &        &                &                          &  \\
\hline
\end {tabular}
\end {center}
\end {table*}

The scatter in the observed abundance ratios are an indication of the efficiency of mixing in the ISM in the era before the formation of the oldest stars we can observe today. The results so far are contradictory.

In Paper V, we demonstrated that the [$\alpha$/Fe] ratios in the EMP giant stars of our sample exhibit very little scatter beyond the observational uncertainty. 
The great uniformity in the [$\alpha$/Fe] ratios of metal-poor stars has recently been demonstrated in more limited samples of turnoff (dwarf) stars also by \citet{Cohen04}, \citet{Arn05}, and \citet{Spite05} and will be further discussed in the next paper of this series (Bonifacio et al., in preparation).
These results are clearly inconsistent with current inhomogeneous chemical evolution models, which predict a scatter of order 1 dex for such elements \citep{Arg02}.

As emphasized by \citet{Arg02}, the initial scatter of a given element ratio,
[X/Fe], is determined by the adopted nucleosynthesis yields. The details of the
chemical evolution model will then determine how fast a homogeneous ISM is
achieved through mixing of the enriched regions. The results of Paper V
indicated that, in order to reproduce the observed low scatter in [$\alpha$/Fe],
the galactic chemical evolution model must employ yields of [$\alpha$/Fe] with little or no dependence on the mass of the progenitor. In homogeneous chemical evolution models \citep{Fra04}, instantaneous mixing is assumed and more variation in the yield can be allowed, because it is integrated over the different stellar masses as the galaxy evolves. 

As the number of EMP stars with high-resolution, high S/N spectroscopy has increased, our ability to quantify the trends and scatter about such
trends for individual elements has improved dramatically as well. In such studies, it is particularly important to use data sets that are reduced 
and analysed in as homogeneous a manner as possible, so as to minimise the
influence of spurious ``observer'' scatter on the behaviours that one seeks to
understand. It was a key goal of our project to produce such data sets.

Thus, Table \ref{tab-scatter} presents estimates of the observed scatter 
of the elemental ratios reported here, following the order of the figures presenting the information, but based exclusively on the stars 
analysed by ourselves.  The first two columns in the table present the
ordinates and abscissae corresponding to each of the figures.  The
number of stars considered in each range of abscissae listed in the 
table appears in the third column, while the range in the parameter under
discussion is listed in the fourth column.

In order to obtain robust estimates of scatter, we must first de-trend
the distributions of the observed ratios.  This is accomplished by
determination of robust locally weighted regression lines ({\it
loess} lines), as described by Cleveland (1979, 1994).  Such lines
have been used before in similar scatter analyses (see, e.g., Ryan et
al.  1996; Carretta et al.  2002).  The scatter about these lines is
then estimated by application of the biweight estimator of scale,
$S_{BI}$, described by Beers, Flynn, \& Gebhardt (1990) \footnote{The
scale matches the dispersion for a normal distribution.}.  

The first entry in the last column of Table \ref{tab-scatter} lists this
estimate.  The quantities in parentheses in this column are the
$1-\sigma$ confidence intervals on this estimate of scatter, obtained
from analysis of 1000 bootstrap resamplings of the data in each of the
given ranges.  In this listing, CL represents the lower interval on
the value of the scatter, while UL represents the upper interval.
These errors are useful for assessing the significance of the
difference between the scales of the data from one range to another.

[Sr/Fe], [Y/Fe], and [Zr/Fe] shows a similar increase of the scatter as 
the metallicity decreases, with a more pronounced effect for Sr. The mean 
ratio  [$<$Sr+Y+Zr$>$/Fe] shows the same behaviour with a lower amplitude. 

Large scatter is also seen in Ba and La, but its variation as a function of metallicity differs from the lighter elements. The dispersion found for Ba seems independent of metallicity, whereas the scatter of La appears much smaller for the most metal-poor stars. Ce, Pr and Nd show much smaller scatter again, in particular Ce for which we measure a bi-weight estimator of only 0.038 dex for the whole sample.  Pr and Nd behave like La with smaller scatter for the most 
metal-poor stars. 

Eu shows a rather high scatter, decreasing as the metallicity decreases. In contrast, Gd, Dy and Er follow the same behaviour as Sr, i.e. an increase in scatter as the metallicity decreases. 

If we now consider the ratios [Eu/Ba] and [Eu/La] as a function of [Fe/H],  
the scatter is smaller by almost an order of magnitude, confirming the common origin of these elements. It is also noteworthy that the scatter is even smaller for the most metal poor metallicity bin.

\subsection {Abundance scatter and inhomogeneous models of galactic 
chemical evolution}

The apparently contradictory abundance results for the $\alpha$- and various
neutron-capture elements in VMP and EMP stars might be reconciled if the sites of significant $r$-process production were diverse and (some of them) rare. 
And we caution that r-II stars {\it are} rare: \citet{Bar05} estimate that they constitute roughly 5\% of the giants with [Fe/H] $< -2.0$. The lower probability of finding them at metallicities below [Fe/H] $\simeq -3.20$ may introduce an artificial decrease of the observed scatter. 

The highly $r$-enriched (r-II) stars have all been found in a very narrow range around [Fe/H] = $-2.9$  \citep{Bar05}. Do we see the onset of a new process at this metallicity? Does this metallicity correspond to the typical metallicity of the building blocks of the halo, originating from systems of similar size (i.e. about the same metallicity) but with different chemical histories (IMF, fraction
of peculiar supernovae), leading to a spread of [n-capture/Fe] but keeping an
$r$-process signature. Did the stars with [Fe/H] $< -3.5$ form out of matter polluted by massive Pop III stars, which could mean that they are
pre-galactic?

It is interesting to note how difficult it has been to find true UMP stars, 
i.e. stars with [Fe/H] $< -4$; in fact, only three are currently known 
\citep{Chi02,Fre05,Nor07}. In a standard closed-box model \citep{Fra90}, we 
would expect to have found several more, if the IMF did not change substantially over time; however, the preferred scheme for the halo formation is an open model, where infall is invoked to explain this ``UMP desert'' \citep{Chi97}.

In the context of an inhomogeneous model of chemical evolution of the Galaxy
\citep{Arg02}, simulations show that the density of stars at [Fe/H] = $-3.0$
and [Fe/H] = $-4.0$ is of the same order \citep[ see their Fig. 7]{Arg02}. As a
consequence, the paucity of UMP stars would require rather fine tuning of
the mixing of supernovae ejecta into the ISM. However, \cite{Kar06} has 
suggested that, alternatively, the absence of UMP stars could be explained with a galactic chemical evolution model where star formation was low or delayed 
for a period after the formation and demise of the first generation of stars, 
due to heating of the ISM by their supernova explosions.

Another possibility is that stars in the inner and outer regions of the halo 
of the Milky Way may have rather different metallicity distribution functions (MDFs). From a kinematic analysis of a local sample of stars from the Sloan 
Digital Sky Survey, \citet{Car07} argue that just such a dichotomy exists, 
with the MDF of the inner-halo stars peaking around [Fe/H] = $-1.6$, that of 
the outer halo around [Fe/H] = $-2.2$. 

The magnitude-limited objective-prism surveys that have identified the most 
metal-poor halo stars to date may thus have been dominated by inner-halo 
objects. If so, simple models of Galactic chemical evolution that match 
the MDFs derived from such surveys may not provide adequate explanations for 
the formation of the Milky Way halo, nor for the detailed chemical composition 
of its most primitive stars.

\section {Conclusions}

This paper has presented accurate, homogeneous abundance determinations for 16 
neutron-capture elements in a sample of 32 VMP and EMP giant stars, for which 
abundances of the lighter elements have been determined earlier (Paper V).
Our data confirm and refine the general results of earlier studies of the neutron-capture elements in EMP stars, and extend them to lower metallicities. 
In particular, the sample of stars below [Fe/H] = $-2.8$ is increased 
significantly.

Our data show the [n-capture/Fe] ratios, and their scatter around the mean 
value, to reach a maximum around [Fe/H] $\simeq -3.0$. Below [Fe/H] 
$\simeq -3.2$, we do not find stars with large overabundances of 
neutron-capture elements relative to the solar ratio. We note, however, 
that the large ``snapshot'' sample of \citet{Bar05} does identify at
least a few stars below [Fe/H] = $-3.0$ with high [Sr/Fe], [Zr/Fe], or 
[Eu/Fe], so a larger sample of accurate data may be needed for a firm 
conclusion.

Adopting Ba as a reference element in the abundance ratios reveals very tight 
anti-correlations of [Sr/Ba], [Y/Ba] and [Zr/Ba] ratios with [Ba/H] abundance from [Ba/H] $\simeq -1.5$ down to [Ba/H] $\simeq -4.5$.These results confirm the need for a second neutron-capture process for the synthesis of the first-peak elements, called the ``weak'' $r$-process \citep{Bus99,Qia00,Wan01}, LEPP process \citep{ Tra04}, or CPR process \citep{Qia07}, or even an entirely new nucleosynthesis mechanism in massive, metal-poor stars ($\nu$p-process, \citet{Fro06}. By subtracting the contributions of the main 
$r$-process, we show that this mechanism is responsible for 90-95\% of the 
amounts of Sr, Y, and Zr in stars with [Ba/H] $> -4.5$. Below this value,
the [Sr/Ba], [Y/Ba], and [Zr/Ba] ratios seem to return to the solar ratio, although the number of stars in this range is small.

As found earlier \citep{Rya96,Mcw98,Hon04}, the [n-capture/Fe] ratios exhibit a much larger dispersion than can be attributed to observational errors, although the scatter in their $[\alpha/Fe]$ and [Fe-peak/Fe] ratios as functions of [Fe/H] is very small. We discuss the implications of these apparently contradictory results on the efficiency of mixing of the primitive ISM in terms of homogeneous vs. inhomogeneous models of galactic chemical evolution.

\begin{acknowledgements}
We thank the ESO staff for assistance during all the runs of our Large 
Programme.  R.C., P.F., V.H., B.P., F.S. \& M.S. thank the PNPS and the PNG for 
their support.  PB and PM acknowledge support from the MIUR/PRIN 2004025729\_002 
and PB from EU contract MEXT-CT-2004-014265 (CIFIST). T.C.B. acknowledges 
partial funding for this work from grants AST 00-98508, AST 00-98549, and AST 
04-06784 as well as from grant PHY 02-16783: Physics Frontiers Center/Joint 
Institute for Nuclear Astrophysics (JINA), all awarded by the U.S. National Science Foundation.  BN and JA thank the Carlsberg Foundation and the Swedish and Danish Natural Science Research Councils for partial financial support of this research.

\end{acknowledgements}

\bibliographystyle{aa}

\end{document}